\documentclass[twoside,12pt]{article}
\usepackage{amssymb}
\def\R{{\mathbb R}}
\def\N{{\mathbb N}}
\def\C{{\Bbb C}}
\def\supp{{\mbox{\rm supp}\, }}
\def\C{~\hbox{\vrule width 0.6pt height 8pt depth 0pt \hskip -3.5pt}C}

\def\kasten{$~~\mbox{\hfil\vrule height6pt width5pt depth-1pt}$ }
\newtheorem{theorem}{Theorem}[section]
\newtheorem{axioms}[theorem]{Axioms}
\newtheorem{proposition}[theorem]{Proposition}
\newtheorem{corollary}[theorem]{Corollary}
\newtheorem{definition}[theorem]{Definition}
\newtheorem{lemma}[theorem]{Lemma}
\newtheorem{remark}[theorem]{Remark}
\begin{document}
\pagestyle{myheadings}
\markboth{H. Gottschalk}{Complex velocity transformations on Krein spaces}
 \thispagestyle{empty}
\begin{flushleft}
\noindent {\Large \bf Complex velocity transformations and the \linebreak
Bisognano--Wichmann theorem for quantum fields acting on Krein spaces}
\end{flushleft}

\

\noindent Hanno Gottschalk\\
{\small Institut f\"ur angewandte Mathematik, \\ Rheinische
Friedrich-Wilhelms-Universit\"at,\\ Wegelerstr. 6, D-53115
Bonn, Germany\\ e-mail: gottscha@wiener.iam.uni-bonn.de}

\

{\noindent \small {\bf Abstract.} It is proven that in indefinite metric quantum field theory
there exists a dense set of analytic vectors for the generator of the one parameter group of $x^0$--$x^1$
velocity transformations. This makes it possible to define complex velocity transformations also for the
indefinite metric case. In combination with the results of Bros -- Epstein -- Moschella \cite{BEM}, proving Bisognano--Wichmann (BW) analyticity
within the linear program, one then obtains a suitable generalization of the BW theorem for local, relativistic quantum fields acting on Krein spaces ("quantum fields with indefinite metric"). 
}

\vspace{.25cm}
{\small \noindent {\bf Keywords}: {\it Quantum fields with indefinite metric, Complex velocity transformations,
Bisognano-Wichmann theorem}}
 \section{Introduction}
The connection of the Unruh effect \cite{Un} describing the Hawking radiation seen by an uniformly accelerated observer
in local quantum field theory (QFT) with the Tomita-Takesaki modular theory \cite{BR,Ta} of v. Neumann algebras has been one of the 
most fascinating achievements in the study of the general structure of QFT, cf. \cite{Bo3,Ha} for an overview. In the Wightmanian formulation
of QFT this connection is given by the Bisognano-Wichmann theorem \cite{BW} which, under a few additional assumptions, can then  be extended to
localized v. Neumann algebras associated with the algebras of unbounded Wightman fields, see e.g. \cite{BWo,BW,In}.

The present work attempts to adapt the Bisognano-Wichmann theorem \cite{BW} to the case of Quantum fields with indefinite metric in the
framework of Morchio and Strocchi \cite{MS,Str1,Str2}. That this should in fact be possible was proven by J. Bros, H. Epstein and U. Moschella in \cite{BEM}. In this reference
the authors show that the KMS condition for the algebra localized in the right wedge ${\cal W}_R=\{x\in\R^d:x^1>|x^0|\}$ 
w.r.t. the one parameter Group of Lorentz boosts in the $x^0$ -- $x^1$ direction follows from the analyticity
properties of Wightman functions from solely the linear program, i.e. without
using positivity. In contrast to this, the original proof of \cite{BW} uses functional calculus for the generators of the Lorentz boosts, and thus positivity, to establish that property.

Quantum fields with indefinite metric have been introduced in the Wightmanian formulation of QFT
in order to deal also with gauge field in local and covariant gauges. Such fields can not be fields with positive metric \cite{Str3,Str2,StrW}. Quantum fields with indefinite metric surely
have to be considered to be artificial and "non-physical" (at least as long as the related physical Hilbert space with positive metric and the algebra of Gauge invariant fields acting on that space 
have not been constructed), but still some motivation can be found, to consider the
BW theorem also in this framework:

\begin{itemize}
\item QFT with indefinite metric is close to \underline{perturbation theory} where local and covariant gauges are needed for technical reasons, see \cite{Ste2,Str2};
\item There are \underline{non-trivial examples} in arbitrary space-time dimension \cite{AG,AGW1,AGW2,To,Os,Ste1,Ste2};
\item The BW theorem might be important to develop a \underline{bounded operator}\linebreak \underline{approach} for quantum fields with indefinite metric defining
wedge algebras as weak commutants in the style of \cite{BW,In};
\item The BW theorem can be seen as a first step towards studying the influence of \underline{constraints} (gauge) to \underline{wedge-duality};
\item It is an interesting technical exercise to develop a suitable \underline{substitute} for \underline{functional calculus} in the case of quantum fields
with indefinite metric.
\end{itemize}
 
This paper is organized as follows: In Section 2 we collect the basic tools for the formulation of QFT with indefinite metric. 

In Section 3 we prove 
a crucial continuity property for GNS-like representations on maximal inner product spaces (Krein spaces \cite{An,Bon}) and we recall basic properties
of the field algebra as the Reeh-Schlieder property. As the vacuum plays a special r\^ole in modular theory, we clarify the status of the vacuum
as the only translation invariant state for QFTs with indefinite metric and a mass gap -- this then also implies the irreducibility of the field algebra 
(see also \cite{MS}). In particular we show that the occurrence of theta-vacua, in the description of G. Morchio and F. Strocchi \cite{MS}, and reducibility of the field algebra due to
classical long range fluctuations \cite{Bu1} are strictly linked
to mass zero phenomena, as of course suggested by these references. 

In Section 4 we define the complex velocity transformations for the case of quantum fields with indefinite metric using
the continuity properties established in Section 3 in connection with functional calculus on the Borchers algebra. Rather than using functional calculus
for $\eta$-symmetric operators, see e.g. \cite{An,HL} and references therein, we follow a "pedestian's approach" and prove the existence of a dense set of entire
analytic vectors for the generator of the Lorentz boosts.

In Section 5 we recall the result of Bisognano--Wichmann analyticity within the linear program as proven by Bros -- Epstein -- Moschella \cite{BEM} and derive a version of
the BW theorem for quantum fields with indefinite metric. In particular, we identify the "modular objects": The anti-unitary implementation of the reflection $\Theta_{0,1}:(x^0,x^1,x^2\ldots,x^{d-1})\to(-x^0,-x^1,x^2,\ldots ,x^{d-1})$ is the "modular conjugation" and 
velocity transformations at imaginary boost parameter $i\pi$ is Tomita's modular operator $\Delta^{1/2}$. For space-time dimension $d=4$ the modular conjugation thus coincides with the rotation around $e_1=(0,1,0,\ldots, 0)$ 
by an angle $\pi$ times the PCT-operator $\Theta:x\to-x$, which is the description given in \cite{BW}. Here, of course, no answer is given to what extent Tomita--Takesaki theory can be generalized to $\eta$- v.Neumann algebras acting on Krein spaces and thus
to what extent the "modular objects" defined here really can be related to some extension of Tomita--Takesaki theory. This might however be an interesting question for the future. At least Tomita--Takesaki theory on the Pontryagin
space $\Pi_1$ has been established in \cite{KS}.

\section{QFT with indefinite metric}
 
Here some notation is introduced and basic results on the GNS-construction on maximal inner product spaces (Krein spaces) are recalled, see however
\cite{Ho2,MS,Yn} for more details. We only consider Bosonic, scalar and chargeless
  QFT's over a $d$--dimensional
Minkowski space-time $(\R^d,\cdot )$. Generalization to arbitrary spin, charge and statistics is straight forward. The associated Borchers algebra \cite{Bo1,Bo2} is the free, unital, involutive, topological
tensor algebra over ${\cal S}_1={\cal S}(\R^d,\C)$, the space of complex Schwartz test functions over $\R^d$.
In more explicit terms, we set
\begin{equation}
\label{2.1eqa}
\underline{\cal S}=\bigoplus_{n=0}^\infty {\cal S}_n,~~~{\cal S}_n={\cal S}(\R^{dn},\C),~{\cal S}_0=\C,
\end{equation}
with $\underline{f}+\underline{h}=(f_0+h_0,f_1+h_1,\ldots)$, $\underline{f}\otimes\underline{h}$ given by $(\underline{f}\otimes\underline{h})_n= \sum_{j,l=0\atop j+l=n}^\infty f_j\otimes h_l$ $\forall n\in\N_0$, ${\bf 1}=(1,0,\ldots)$ and
$\underline{f}^*=(f_0^*,\ldots,f_n^*,0\ldots)$ with $f_l^*(x_1,\ldots,x_l)=\overline{f_l(x_l,\ldots,x_1)}$. Here the bar denotes complex conjugation and the product always exists as for $\underline{f}\in\underline{\cal S}$
only finitely many $f_l\in{\cal S}_l$ are different from zero.

The canonic action $\alpha:P^\uparrow_+\to\mbox{Aut}({\cal S}_l)$ of the orthochronous, proper Poincar\'e group on ${\cal S}_l$, $l\in\N$ induces
a representation $\underline{\alpha}:P^\uparrow_+\to\mbox{Aut}(\underline{\cal S})$ by continuous $*$-algebra homomorphisms on $\underline{\cal S}$.

The spectral ideal $I_{\rm sp}\subset \underline{\cal S}$ is the left-ideal generated by elements of the form
\begin{equation}
\label{2.2eqa}
\underline{g}=\underline{g}(\underline{f},h)=\int_{\R^d}\underline{\alpha}_{\{1,a\}}(\underline{f})\, h(a) \, da\in\underline{\cal S}
\end{equation}
for $\supp\hat h\cap \bar V_0^+=\emptyset$ where $h\in{\cal S}_1$, $\hat h(k)=\int_{\R^d}e^{ik\cdot a}h(a)\, da$ and
$\bar V_0^+=\{ k\in\R^d, k^0\geq 0, k\cdot k\geq 0\}$ is the closed forward light cone. The related left-ideal $I_{\rm sp}^{m}$ where
in the definition of $I_{sp}$ $\bar V_0^+$ is replaced by $\{0\}\cup\bar V_m^{+}=\{0\}\cup \{ k\in\R^d, k^0\geq 0, k\cdot k\geq m^2\}$, $m>0$, is called
spectral ideal with mass-gap $m$.

A further useful ideal in $\underline{\cal S}$ is the two-sided ideal generated by $(0,0,[f_1,h_1],0,\linebreak \ldots)$ with $f_1,h_1\in{\cal S}_1$ and $\mbox{supp}~f_1$, $ \mbox{supp}~h_1$ space-like separated,
i.e. $(x-y)\cdot(x-y)<0$ $\forall x\in\supp~ f_1, y\in\supp~ h_1$. It is called the ideal of locality.

We recall that $\underline{\cal S}$ is endowed with the strongest topology, such that the relative topology 
of ${\cal S}_n$ in $ \underline{\cal S}$ is the Schwartz topology (direct sum topology).
Let $\underline{\cal S}'=\underline{\cal S}'(\R^d,\C)$ be the topological dual space of $\underline{\cal S}$.
Then $\underline{R}\in\underline{\cal S}'$ is of the form $\underline{R}=(R_0,R_1,R_2,\ldots)$ with
$R_0\in \C, R_n\in {\cal }S_n', n\in\N$. Furthermore, any such sequence defines uniquely an 
element of $\underline{\cal S}'$. An element $\underline{W}\in\underline{\cal S}'$ is called a Wightman functional
if it fulfills the following set of conditions, which are also called the modified Wightman axioms, cf. \cite{MS,Yn}:

\begin{axioms}
\label{2.1ax}
\rm
A1) Temperedness and normalization: $\underline{W}\in\underline{\cal S}'$ and $W_0=1$.

\noindent A2) Poincar\'e invariance: $\underline{W}( \underline{\alpha}_{\{\Lambda,a\}}(\underline{f}))=\underline{W}(\underline{f})~ \forall \{\Lambda,a\}\in \tilde P^\uparrow_+$, $\underline{f}\in\underline{\cal S}$.

\noindent A3) Spectral property: Let $ I_{\rm sp}$ be the spectral (left) ideal in $\underline{\cal S}$.
Then $I_{\rm sp}\subseteq \mbox{kernel}~\underline{W}$.

\noindent A4) Locality: Let $ I_{\rm loc}$ be the (two-sided) ideal of locality in $ \underline{ \cal S}$.
Then $I_{\rm loc}\subseteq \mbox{kernel}~\underline{W}$.

\noindent A5) Hilbert space structure condition (HSSC): There exists a Hilbert seminorm $\underline{p}$ on $\underline{\cal S}$ s.t. $\left|\underline{W}(\underline{f}^*\otimes \underline{g})\right|\leq \underline{p}(\underline{f})\underline{p}(\underline{g}) \forall \underline{f},\underline{g} \in \underline{S}$.

\noindent A6) Cluster Property: $\lim_{t\to \infty}\underline{W}(\underline{f}\otimes \underline{\alpha }_{\{1,ta\}}(\underline {g})) = \underline{W}(\underline{f})~\underline{W}(\underline{h})$ $\forall \underline{f},\underline{g}\in \underline{S}, a\in \R^{d} $ space like.

\noindent A7) Hermiticity: $\underline{W}(\underline{f}^*)=\overline{\underline{W}(\underline{f})}$ $\forall \underline{f}\in\underline{\cal S}$.
\end{axioms}

We say that the Hilbert seminorm $\underline{p}$ in A5) is of Sobolev type, if
\begin{equation}
\label{2.3eqa}
\underline{p}(\underline{f})=\sum_{l=0}^\infty p_n(f_n)~~\forall\underline{f}\in\underline{\cal S}
\end{equation} s.t. $p_n$ defined on
${\cal S}_n$ up to multiplication with a positive constant is given by a Hilbert norm of Sobolev type \cite{GV}, 
\begin{equation}
\label{2.3aeqa}
p_n(f_n)^2= c^2\int_{\R^{dn}}\left|\prod_{l=1}^n(1+|x_l|^2)^{N/2}(1-\Delta_{x_l})^{L/2}f_n(x_1,\ldots,x_n)\right|^2dx_1\cdots dx_n,
\end{equation} 
for some possibly $n$-dependent $L,N\in\N_0$ and $c>0$ with $\Delta_{x_l}=\sum_{l=1}^n\partial^2/(\partial x_l)^2$. $\underline{p}$ is called Sobolev dominated, if there exists a Hilbert norm $\underline{p}'$ of Sobolev type s.t. $\underline{p}\leq\underline{p}'$ or,
in other words, if $\underline{p}$ is continuous w.r.t. the $\underline{\cal S}$-topology.

If $I_{\rm sp}$ in A3) can be replaced with some bigger $I_{\rm sp}^m$, $m>0$, then we say that $\underline{W}$ fulfills the strong spectral condition.
One can show by explicit calculations that the axioms A1)-A4),A6) and A7) are equivalent to the usual Wightman axioms \cite{SW}, whereas positivity has been
replaced by the weaker assumption A5). Nevertheless, A5) is enough to get an analogue of the Wightman reconstruction theorem on maximal, non degenerate
inner product spaces as follows:

A metric operator $\eta:{\cal H}\to{\cal H}$ by definition is a self adjoint operator
on a complex separable Hilbert space $({\cal H},(.,.))$ with $\eta^2=1$. $\eta$ induces a second,
in general indefinite, inner product on ${\cal H}$ via $\langle.,.\rangle=(.,\eta.)$ and $\eta$ then 
gives a fundamental decomposition of ${\cal H}$ into intrinsically complete subspaces ${\cal H}={\cal H}^+\oplus{\cal H}^-$
s.t. $\langle.,.\rangle$ restricted to ${\cal H}^\pm$ is positive/negative definite. Hence $({\cal H},\langle.,.\rangle)$ is a Krein space
\cite{An,Bon}. It should be mentioned that $(.,.)$, $\eta$ and the fundamental decomposition ${\cal H}^\pm$ are not unique given 
$({\cal H},\langle.,.\rangle)$; $\eta$ for example can be replaced by $\eta_C={\rm sign}(|C|^{-2}\eta) $ for a bounded, continuously invertible
operator on ${\cal H}$ leading to a in general different, but topologically equivalent, $(.,.)_C=(.,|C|^{2}||C|^{-2}\eta|.)$
and also a new fundamental decomposition according to $\eta_C$. 

Let  ${\cal D}$ be a dense and linear subspace of ${\cal H}$. The set of
 (possibly unbounded) Hilbert space operators ${\sf L}:{\cal D}\to{\cal D}$
  with (restricted) $\eta$-adjoint ${\sf L}^{[*]}=\eta {\sf L}^*\eta|_{\cal D}:{\cal D}\to{\cal D}$
   is denoted with ${\sf O}_\eta ({\cal D})$. Clearly, ${\sf O}_\eta ({\cal D})$ is an unital algebra
    with involution $[*]$. The canonical topology on ${\sf O}_\eta ({\cal D})$ is generated
	 by the seminorms ${\sf L}\to|\langle\Psi_1, {\sf L} \Psi_2\rangle|,~\Psi_1,\Psi_2\in {\cal D}$. We then
	  have the following theorem:

\begin{theorem}
\label{2.1theo}
Let $\underline{W}\in\underline{\cal S}$ be a Wightman functional which fulfills the Axioms \ref{2.1ax}. Then

\noindent (i) There is a Hilbert space $({\cal H},(.,.))$ with a distinguished normalized vector $\Omega\in {\cal H}$ called the vacuum, a metric operator $\eta$ with $\eta\Omega=\Omega$ inducing a nondegenerate inner product $\langle.,.\rangle=(.,\eta .)$ and a continuous $*$-algebra representation $\phi:\underline{\cal S}\to {\sf O}_\eta ({\cal D})$ with ${\cal D}=\phi(\underline{S})\Omega$ which is connected to the Wightman functional $\underline{W}$ via $\underline{W}(\underline{f})=\langle \Omega,\phi(\underline{f})\Omega\rangle\forall \underline{f}\in \underline{\cal S}$.

\noindent (ii) There is a $\eta$-unitary continuous representation ${\sf U}: \tilde P_+^\uparrow \to{\sf O}_\eta ({\cal D})$ ($\, {\sf U}^{[*]}={\sf U}^{-1}$) such that ${\sf U}( \Lambda,a) \phi (\underline{f}){\sf U}(\Lambda,a)^{-1}=\phi(\underline{\alpha}_{\{\Lambda,a\}}(\underline{f}))~\forall \underline{f}\in\underline{\cal S}, \{\Lambda,a\}\in \tilde P_+^\uparrow$ and $\Omega$ is invariant under the action of ${\sf U}$.

\noindent (iii) $\phi$ fulfills the spectral condition $\phi(I_{\rm sp})\Omega=0$.

\noindent (iv) $\phi$ is a local representation in the sense that $I_{\rm loc} \subseteq \mbox{\rm kernel}~\phi$.

\noindent (v) For $\Psi_1,\Psi_2\in{\cal D}$ and $a\in \R^d$ space like, we get $\lim_{t\to\infty}\langle\Psi_1,{\sf U}(1,ta)\Psi_2\rangle=\langle\Psi_1,\Omega\rangle\langle\Omega,\Psi_2\rangle$.

A quadruple $(({\cal H},\langle.,.\rangle,\Omega),\eta,{\sf U},\phi)$ is called a local relativistic QFT in indefinite metric.

Conversely, let $(({\cal H},\langle.,.\rangle,\Omega),\eta,{\sf U},\phi)$ be a local relativistic QFT in indefinite metric. Then
$ \underline{W}(\underline{f})=\langle \Omega,\phi(\underline{f})\Omega\rangle  ~ \forall \underline{f}\in\underline{\cal S}$ defines a Wightman functional $\underline{W}\in\underline{\cal S}'$ which fulfills the Axioms \ref{2.1ax}.
\end{theorem}
As we will come back to  some details of the proof of Theorem \ref{2.1theo} in the next section, we recall
these points here while for the rest of the proof we refer to \cite{MS,Yn,Ho2}.

\noindent {\bf Sketch of Proof.} Without loss of generality one can assume that $\underline{p}$ is a Hilbert norm. If $\underline{p}$ is only a Hilbert seminorm we can replace it
by the Hilbert norm $\underline{p}'=\sqrt{\underline{p}^2+\underline{p}_1^2}$ where $\underline{p}_1$ is a Hilbert norm on $\underline{\cal S}$ which not necessarily dominates $\underline{W}$. Such $\underline{p}_1$ can e.g. be chosen as a
direct sum of Sobolev norms on ${\cal S}_n$.  

Then, ${\cal H}_1=\overline{\underline{\cal S}}^{\underline{p}}$ defines the Hilbert space completion of ${\cal D}_1=\underline{\cal S}$ w.r.t. $\underline{p}$. The Hermitian inner product $\langle \underline{f},\underline{g}\rangle=\underline{W}(\underline{f}^*\otimes\underline{g})$ defined
for $\underline{f},\underline{g}\in{\cal D}_1$ is continuous and thus extends uniquely to ${\cal H}_1$. By the Riesz representation theorem there exists a self adjoint operator $\eta_1$ bounded by one s.t. $\langle.,.\rangle=(.,\eta_1.)_1$.
The algebra $\underline{\cal S}$ acts via the identical representation $\phi_1$ by multiplication from the left on ${\cal D}_1\subseteq {\cal H}_1$. Likewise, a $\eta_1$-unitary 
continuous, representation ${\sf U}_1:P^\uparrow_+\to{\sf O}_{\eta_1}({\cal D}_1)$ is defined by the action of $\underline{\alpha}$ on ${\cal D}_1$.

Let ${\cal K}_0=\{ v\in{\cal H}_1:\langle w,v\rangle=0\, \forall w\in{\cal H}_1\}$ and $\underline{\cal S}_0=\{ \underline{f}:\underline{W}(\underline{g}\otimes\underline{f})=0$ $\forall g\in\underline{\cal S}\}$.
Then, ${\cal K}_0$ is a closed subspace of ${\cal H}_1$. Let ${\cal H}_2={\cal H}_1/{\cal K}_0$ be quotient Hilbert space of ${\cal H}_1$ and ${\cal K}_0$ with scalar product $(.,.)_2$. By Hermiticity, linearity and
definition of ${\cal K}_0$ $\langle.,.\rangle$ is also defined on the quotient space ${\cal H}_2$. Let $\pi:{\cal H}_1\to{\cal H}_2$ be the quotient map. Then,
by the definition of ${\cal K}_0$, $(\pi(\underline{f}),\pi(\underline{g}))_2$ still dominates $\langle.,.\rangle$.
Thus, there exists a self-adjoint, operator $\eta_2$ bounded by one s.t. $\langle.,.\rangle=(.,\eta_2.)_2$. Furthermore we set ${\cal D}=\pi({\cal D}_1)\cong \underline{\cal S}/\underline{\cal S}_0$ as ${\cal D}_1\cap {\cal K}_0=
\phi_1({\cal S}_0)$. As $\underline{\cal S}_0$ is a left ideal and is taken into itself by the action of $\underline{\alpha}$, $\phi=\pi\circ\phi_1$ and ${\sf U}=\pi\circ{\sf U}_1$ define a
left-representation of $\underline{\cal S}$ and $P^\uparrow_+$, respectively, in ${\sf O}_{\eta_2}({\cal D})$.

Let $(.,.)=(.,|\eta_2|.)_2$. Then, the metric operator $\eta$ defined by $\langle.,.\rangle=(.,\eta.)$ is $\eta={\rm sign}(\eta_2)$. Thus, also $(.,.)$ dominates $\langle.,.\rangle$ and obviously $\|.\|\leq\|.\|_2$. We can now set
${\cal H}$ to be the completion of ${\cal D}$ w.r.t. $\|.\|$, extend $\eta$ and $\langle.,.\rangle$ to ${\cal H}$ in order to obtain a Krein space representation, as described by the assertion of the theorem. The representations $\phi$ and
${\sf U}$ in ${\sf O}_\eta({\cal D})$ then fulfill the requirements (i)-(v), as proven in \cite{MS,Ho2,SW,Yn} (see also \cite{Do}). For the fact that $\eta$ can be chosen s.t. $\eta\Omega=\Omega$, $\Omega=\pi(1,0,\ldots)$, cf. \cite{Ho2}. \kasten

Let $\Theta_{0,1}$ be the anti-linear PCT-like transformation given by the action\linebreak
 $\alpha_{\{\Theta_{0,1},0\}}(f_n)(x_1,\ldots,x_n)=\overline{f_n(\Theta_{0,1}x_1,\ldots,\Theta_{0,1}x_n)}$ for $f_n\in{\cal S}_n$, $\Theta_{0,1}x$ as in the introduction, and $\underline{\alpha}_{\{\Theta_{0,1},0\}}$ the related action on $\underline{\cal S}$.
From the proof of Theorem \ref{2.1theo} one immediately gets (cf. \cite{Str2})
\begin{corollary}
\label{2.1cor}
Let $\underline{p}$ in A5) be such that $\underline{p}(\underline{\alpha}_{\{\Lambda,0\}}(\underline{f}))=\underline{p}(\underline{f})$ $\forall \underline{f}\in\underline{\cal S}$ and $\Lambda\in G\subseteq L$, with $G$ generated by
rotations (in a given frame of reference) and the PCT-like transformation $\Theta_{0,1}$. Then, ${\sf U}$ extends to all $G$ and ${\sf U}(G)$ commutes with the metric operator $\eta$ constructed in the Proof of Theorem \ref{2.1theo}. In particular, ${\sf U}(G)$
is represented by (anti-) unitary, and hence bounded, operators on ${\cal H}$.  
\end{corollary}
\noindent{\bf Proof.} By the PCT-theorem in $d$ space-time dimensions, which does not require positivity (here we assumed the standard connection of spin and statistics, which is required for the PCT-theorem -- in theories with indefinite metric where also
ghost fields are included, the results of this article on the BW theorem do not apply), $\underline{W}$ is invariant under the anti-linear PCT-like transformation $\Theta_{0,1}$ \cite{BEM,SW} and thus ${\sf U}(G)$ ist (anti-)$\eta$-unitarily represented. As $\underline{p}$ is invariant under the action of $G$,
$\eta_1$ commutes with ${\sf U}_1(G)$. Since the construction of $\cal H$ consists of twice changing the operators $\eta_1$ and $\eta_2$ by functions of these operators, ${\sf U}(G)$ also commutes with $\eta$. (Anti-) unitarity of ${\sf U}(G)$ now follows
from $\eta^2=1$.\kasten 

It is demonstrated in \cite{MS2} Theorem 3.1 that Sobolev dominated Hilbert norms which do not fulfill the condition of Corollary \ref{2.1cor} can be replaced by equivalent norms which fulfill that condition. It is automatically fulfilled by Hilbert norms of Sobolev type.

To close this section we would like to address two points concerning the usefulness of Theorem \ref{2.1theo} to describe the physics of quantum fields with indefinite metric. 
The first point is that the construction of ${\cal H}$ depends
not only on $\underline{W}$ but possibly depends also on the auxilary Hilbert seminorm $\underline{p}$. In fact, unless the fundamental decomposition ${\cal H}^\pm$ of $\cal H$ given by $\eta$ has either ${\cal H}^+$ or ${\cal H}^-$ 
intrinsically complete, i.e. complete w.r.t. the topology generated by the restriction of $\langle.,.\rangle$ to ${\cal H}^\pm$,
recent results prove that there are infinitely many topologically inequivalent maximal Hilbert space structures \cite{CG,Ho3,HoX}. In the generic
situation in QFT the fundamental decomposition ${\cal H}^\pm$ can not be expected to have an intrinsically complete component. On the other hand it is argued in \cite{MS1,MS2} that
exactly this non-uniqueness can be exploited 
for the  construction of charged states in certain infra-red representations (depending on the space-time splitting) and thus is a necessary feature of gauge QFT. Another application involving an adequate choice
of the closure of local states is e.g. Bosonization in low-dimensional QFT, cf. \cite{Str2}. 
The present article deals with this issue in the way
that we prove our results for {\it all} such maximal Hilbert space structures which originate from Sobolev dominated $\underline{p}$.

The second point, which has recently been expressed by O. Steinmann, is about that maximal Hilbert topologies on the space(s) ${\cal H}$ might have little to do with the topology on the space of physical states, as in the case of quantum electrodynamics it is
not possible to construct charged states for the Maxwell (Gauss) charge in a physical Hilbert space constructed from ${\cal H}$ with the help of an abstract Gupta-Bleuler formalism in the spirit of \cite{MS}, cf. Theorem 7.3 of \cite{Ste2}.
Instead, Steinmann proposes to work with the inner product space ${\cal D}=\underline{\cal S}/\underline{\cal S}_0$ and to avoid a (possibly non-unique) maximal Hilbert closure. Endowing ${\cal D}$ with a suitable quotient topology, the arguments of the present article can also be formulated
without reference to maximal Hilbert space closures of ${\cal D}$, this will in fact be the main strategy of this article. As the results which take reference to Hilbert space structures are mathematically slightly stronger and it might be of interest to compare them with
results on functional calculus on Krein spaces \cite{An,HL}, here we use the language of Krein spaces and do not restrict the analysis to ${\cal D}$.  

\begin{remark}
\label{2.1rem}
{\rm Section 5 of \cite{MS1} states
that charged, physical states {\em can be} constructed in a closure of the local states. Theorem 7.3 of \cite{Ste2} is just the opposite of this statement. 
As far as I can see, this contradiction can be traced back to different assumptions in \cite{MS1} and \cite{Ste2}. In the \cite{Ste2} it is assumed that
 the local states are in the domain of definition of the Maxwell charge $Q$. This assumption leaves open
a kind of loophole as $Q$ resulting from a closure of the Maxwell charge defined on states of the type of eq. (86) of \cite{MS1} might not contain all local states in its domain.

On the other hand, the construction of the metric operator in \cite{MS1} according to
conditions ${\bf A}_2$) for the "in"-field and eq. (70), leading to the construction of charged physical states, could also be problematic: The vacuum expectation values
of local and "in"-fields, which can be calculated from the local theory,  partially fix the action of the metric operator on the "in"-fields, as e.g. sequences of states converging in the "in"-space must have finite inner product with the local states. 
In a hypothetic, nonperturbative QED ${\bf A}_2$) and eq. (70) of \cite{MS1}
might not match with this {\em a priori} requirement on the metric operator.  In a related situation (again not exactly the same as in \cite{MS1}), Steinmann gives evidence from perturbation theory that there could in fact be a problem, cf. Chapter 12.4 of \cite{Ste2}. But the
Hilbert space of an interacting theory usually is singular to the Hilbert space of the free theory, on which perturbation theory is carried out, thus perturbative arguments on the states of the interacting theory are not fully conclusive.  
\kasten}
\end{remark}

 \section{Some properties of the field algebra}
The reason why we went trough the well-known construction in the proof of Theorem \ref{2.1theo} is that the continuity $\phi:\underline{\cal S}\to{\sf O}_\eta({\cal D})$ for most applications
is insufficient, as the topology on ${\sf O}_\eta({\cal D})$ is very weak. From the above construction we however get that
\begin{equation}
\label{3.1eqa}
\|\phi(\underline{f})\Omega\|\leq \underline{p}(\underline{f})~~\forall \underline{f}\in\underline{\cal S}
\end{equation}
and we thus get the following lemma, which gives sufficient continuity properties of the representation $\phi$
to effectively control certain constructions on ${\cal H}$. 
\begin{lemma}
\label{3.1lem}
Suppose $\underline{W}$ fulfills the HSSC A5) with $\underline{p}$ Sobolev dominated. 
Then, $\phi$ is strongly continuous in the sense 
$\underline{f}_n\to \underline{f}$ in $\underline{S}$ $\Rightarrow$ $\phi(\underline{f}_n)\Omega\to\phi(\underline{f})\Omega$ strongly in ${\cal H}$.
\end{lemma}
{\bf Proof.} As $\underline{\cal S}$ carries the direct sum topology, $\underline{f}_n\to\underline{f}$ implies the convergence of any component in ${\cal S}_n$ and the
existence of a maximal $N$ s.t. the components of $\underline{f}_n$ in ${\cal S}_l$, $l>N$, are equal to zero $\forall n\in\N$. As the topology on ${\cal S}_n$ is generated by Sobolev dominated norms and convergence in
${\cal S}_n$ thus implies the convergence in any specific Sobolev norm, we get $0\leq\underline{p}(\underline{f}_n-\underline{f})\leq \underline{p}'(\underline{f}_n-\underline{f})\to 0$, as $n\to\infty$, for $\underline{p}'$ a seminorm of Sobolev type 
dominating $\underline{p}$. Together with (\ref{3.1eqa}) this gives $\|\phi(\underline{f}_n)
\Omega-\phi(\underline{f})\Omega\|=\|\phi(\underline{f}_n-\underline{f})\Omega\|\to 0$. \kasten

From now on we assume that the maximal Hilbert space ${\cal H}$ with metric operator $\eta$ has been constructed from a Sobolev dominated Hilbert norm $\underline{p}$. That such a $\underline{p}$ exists
can be checked explicitly using the following sufficient condition, which has been verified in \cite{AG,AGW2} for models of QFT with indefinite metric and nontrivial scattering matrix. It should be no major problem to verify
the condition of Theorem \ref{3.1theo} also for the truncated Wightman functions  of $n$-th order perturbation theory of \cite{Os,Ste1,Ste2} and the conformally invariant Wightman functions as classified by Nikolov and Todorov, \cite{To}. 
\begin{theorem}
\label{3.1theo}
Let $\|.\|$ be a Schwartz norm on ${\cal S}_1$. If the truncated Wightman functions $W_n^T$ associated with $\underline{W}$ (cf. (\ref{3.2eqa}) below) are continuous\footnote{As ${\cal S}_1$ is a nuclear space, the tensor product of norms exists, cf. e.g. \cite{Ho1}.}
w.r.t. $\|.\|^{\otimes n}$ $\forall n\in\N$ then $\underline{W}$ fulfills the HSSC with respect to a Hilbert norm $\underline{p}$ of Sobolev type.
\end{theorem}
For the proof see \cite{AGW2} and \cite{Ho1}.
 Here we only recall that the Schwarz norms $\|.\|=\|.\|_{L,N}$, $L,N\in\N_0$, on ${\cal S}_1$ are given by
\begin{equation}
\|f\|_{L,N}=\sup_{x\in\R^d,|\beta|\leq L}\left|(1+|x|^2)^{N/2}\partial^{\beta}f(x)\right|,
\end{equation}
where $\beta$ is a multiindex and $\partial^{\beta}=\prod_{l=0}^{d-1}\partial^{|\beta^l|}/(\partial x^l)^{\beta_l}$. The topology generated by the family of Schwartz norms is equivalent to the topology generated by 
the Sobolev norms, which also generate the Schwartz topology. For concrete estimates it is often more convenient to work with Schwartz norms,
therefore Theorem \ref{3.1theo} is formulated for such norms. 

For ${\cal A}\subseteq {\sf O}_\eta({\cal D})$ the weak commutant of ${\cal A}$ is given by
\begin{equation}
\label{5.2eqa}
{\cal A}'=\left\{ C\in {\cal B}({\cal H}):\langle \Psi_1,C{\sf L}\Psi_2\rangle=\langle {\sf L}^{[*]}\Psi_1,C\Psi_2\rangle ~~\forall \Psi_1,\Psi_2\in{\cal D}, {\sf L}\in{\cal A}\right \}.
\end{equation}
Here ${\cal B}({\cal H})$ is the set of bounded operators on ${\cal H}$. ${\cal A}$ is called irreducible if ${\cal A}'=\C 1_{\cal H}$.

For ${\cal O}\subseteq\R^d$ we define $\underline{\cal S}({\cal O})$ to be the unital *-algebra of those $\underline{f}\in\underline{\cal S}$ which have components $f_n$, $\mbox{supp}~f_n\subseteq {\cal O}^{\times n}$. We set ${\cal P}({\cal O})=\phi(\underline{\cal S}({\cal O}))$ and we call this set the polynomial
 field algebra localized in ${\cal O}$. Furthermore, we let ${\sf E}_0=(\Omega,.)\Omega\in{\sf O}_\eta({\cal D})$  be
  the projection operator onto the vacuum. One then gets the Reeh--Schlieder property
   of the field in analogy to the positive metric case (the proof essentially is due to \cite{MS,Str1}):

\begin{theorem}
\label{3.2theo}
(i) For ${\cal O}\subseteq \R^d$ open and not empty, $\Omega$ is cyclic w.r.t. ${\cal P}({\cal O})$;

\noindent (ii)  For ${\cal O}$ as in (i), the set of operators $\{{\cal P}({\cal O}),{\sf E}_0\}$ is irreducible.

\noindent (iii) For ${\cal O}$ as in (i) s.t. ${\cal O}'=\{x\in\R^d, (x-y)\cdot(x-y)<0~\forall y\in {\cal O}\}$ has nonempty
open interior, $\Omega$ is standard (i.e. cyclic and separating) for ${\cal P}({\cal O})$.
\end{theorem}

\noindent {\bf Proof.} (i) By Lemma \ref{3.1lem}, $\underline{\cal S}\ni\underline{f}\to\langle \Psi_1,\phi(\underline{f})\Omega\rangle$ defines a functional in $\underline{\cal S}'$ for arbitrary $\Psi_1\in{\cal H}$. This step has been used also in \cite{Str1,MS} without being explicitly mentioned. For the standard Wightman axioms it follows from positivity, cf. \cite{SW}. As in \cite{SW} we may now conclude that the $n$-point functions associated to such Wightman functionals are boundary values of analytic functions and thus vanish everywhere, if they vanish in ${\cal O}$. Now suppose that $\Psi$ is perpendicular (w.r.t. $(.,.)$) to ${\cal P}({\cal O})\Omega$. This implies for $\Psi_1=\eta \Psi$, using the above argument, that $\langle \Psi_1,\phi(\underline{f})\Omega\rangle=0\forall \underline{f}\in\underline{\cal S}$. Thus $\Psi_1=0$ and $\Psi=\eta\Psi_1=0$.

(ii) We have to show that for any $C\in{\cal B}(\cal H)$
$$
\left\langle \Psi_1,C\phi(\underline{f})\Psi_2\right\rangle=\left\langle\phi(\underline{f}^*)\Psi_1,C\Psi_2\right \rangle\forall \underline{f}\in\underline{\cal S}({\cal O}),\Psi_1,\Psi_2\in{\cal D}
$$
and $C{\sf E}_0={\sf E}_0C$ implies $C=c1_{\cal H}$ for some $c\in\C$. Let $\Psi_1\in{\cal D}$ and $\Psi_2\in {\cal P}({\cal O})\Omega$, i.e. $ \Psi_2=\phi(\underline{f})\Omega$ for some $\underline{f}\in\underline{\cal S}({\cal O})$. We get in analogy to Equation (4-16) of \cite{SW}:
\begin{eqnarray*}
\langle \Psi_1,C\Psi_2\rangle&=& \langle \phi(\underline{f}^*)\Psi_1,C\Omega\rangle
= \langle \phi (\underline{f}^*)\Psi_1,C{\sf E}_0\Omega\rangle\\
&=& ( \phi(\underline{f}^*)\Psi_1,\eta {\sf E}_0C\Omega)=(\phi(\underline{f}^*)\Psi_1,\eta \Omega)(\Omega,C\Omega)\\
&=&\langle \phi(\underline{f}^*)\Psi_1,\Omega\rangle (\Omega,C\Omega)=\langle\Psi_1,\Psi_2\rangle(\Omega,C\Omega).
\end{eqnarray*}
Since ${\cal P}({\cal O})\Omega$ is dense in ${\cal H}$ by (i), this implies $C=(\Omega,C\Omega)1_{\cal H}$.

(iii) Follows from property (i) for ${\cal P}({\cal O})$ and ${\cal P}({\cal O}')$ since Locality implies that these algebras commute and
thus cyclicity of $\Omega$ for ${\cal P}({\cal O}')$ implies separability of $\Omega$ for ${\cal P}({\cal O})$. 
\kasten

We want to adapt further results of positive metric QFT to the case of indefinite metric
 QFT, namely the irreducibility of the field algebra and the uniqueness of the vacuum.
  The proofs in the positive metric case use the spectral resolution of the translation
   group -- something which is not at disposal in ``indefinite metric''.
     Instead we use a smooth cut-off function to separate the vacuum from the remaining 
	 states and we have to restrict ourselves to the case of QFT's with a mass-gap $m>0$.

Since ${\sf U}$ is only densely defined,  one calls $\tilde \Omega\in{\cal H}$ invariant under the action
 of the translation group ${\sf U}(1,a)$ if $\langle\tilde\Omega,{\sf U}(1,a)\Psi_1\rangle=\langle\tilde\Omega,\Psi_1\rangle \forall a\in\R^d,\Psi_1\in {\cal D}$, cf. \cite{Str2}.

\begin{theorem}
\label{3.3theo}
Suppose that $\underline{W}$ fulfills the strong spectral condition with mass gap $m>0$.  Then

\noindent (i) ${\cal P}(\R^d)$ is irreducible;

\noindent (ii) The vacuum $\Omega$ is the unique translation invariant state in ${\cal H}$ (up to multiplication with a constant).
\end{theorem}
Related results have been proven by Morchio and Strocchi \cite{MS}: (ii) has been verified for non-maximal Hilbert spaces of Sobolev type and (i) has been deduced
from the assumption of (essential) uniqueness of the vacuum. Here we extend these results also to maximal Hilbert closures originating from a Sobolev dominated HSSC and
we prove that such closures do not contain any theta-vacua \cite{MS}, provided the strong spectral condition holds. That this statement can not be obtained in the mass-zero case
is proven in \cite{MS} by explicit examples. These examples however do not fulfill the cluster property, which in \cite{MS} is not included to the modified Wightman axioms. 
Such a statement fits nicely into the picture that there is a connection of symmetry breaking and the presence of massless fields \cite{Bu}.  Also, (i) shows that a potential reducibility
of the algebra of local fields can occur in the mass-zero case only. For massless gauge theories reducibility of the representation can be expected resulting 
from classical long range fluctuations, see \cite{Bu1} for the example of QED. For non-trivial
examples of massive theories see \cite{AG}. 

The proof of Theorem \ref{3.3theo} starts with some technical preparations: Let $\underline{f}\in\underline{\cal S}$ and
 $h\in{\cal S}_1$ and $\underline{g}=\underline{g}(\underline{f},h)$ be defined as in (\ref{2.2eqa}).
We assume that ${\rm supp}~\hat f_n$ is compact $\forall n\in\N$ with $\hat f_n$ the Fourier transform of $f_n$ in all arguments,
 $\mbox{supp }\hat h\subseteq (m/3,\infty)\times \R^{d-1}$
  and $\hat h=1$ on the set $\{k=k_1+\cdots+k_n:(k_1,\ldots,k_n)\in\mbox{supp }
  \hat f_n,n\in\N\}\cap \bar V_{m/2}^+$.
   We denote the vector space of all such $\underline{g}$
    which are obtained in this way by $\underline{\cal S}^+$.
	
 Let $\lambda_l=(\lambda^1_l,\ldots,\lambda^j_l)\subseteq (1,\ldots,n)$ where the inclusion means that $\lambda_l$ is a subset of $\{1,\ldots,n\}$ and the natural order of $(1,\ldots,n)$ is preserved. Let ${\cal P}(1,\ldots,n)$ denote the collection of all partitions of $(1,\ldots,n)$ into disjoint sets $\lambda_l$, i.e. for $\lambda\in {\cal P}(1,\ldots,n)$ we have $\lambda=\{\lambda_1,\ldots,\lambda_r\}$ for some $r$ where $\lambda_l\subseteq (1,\ldots,n)$, $\lambda_l\cap\lambda_{l'}=\emptyset$ for $l\not = l'$ and $\cup_{l=1}^r\lambda_l=\{ 1,\ldots,n\}$. Given a Wightman functional $\underline{W}\in\underline{\cal S}'$ and $\lambda_l=(\lambda_l^1,\ldots,\lambda_l^j)$, we set $W(\lambda_l)=W_j(x_{\lambda_l^1},\ldots,x_{\lambda_l^j})$.

With this definition at hand we can recursively define the truncated Wightman functional $\underline{W}^T\in\underline{\cal S}'$ associated to $\underline{W}\in\underline{\cal S}'$ via $W_0^T=0$ and
\begin{equation}
\label{3.2eqa}
W(1,\ldots,n)=\sum_{\lambda\in{\cal P}(1,\ldots,n)}\prod_{l=1}^{|\lambda|} W^T(\lambda_l)~,n\in\N,
\end{equation}
where $|\lambda|$ is the number of sets $\lambda_l$ in $\lambda$. We also recall that the translation invariance of $\underline{W}$, clustering and the strong spectral property imply
${\rm supp}~\hat W_n^T(k_1,\ldots,k_n)\linebreak \subseteq\{(k_1,\ldots,k_n)\in\R^{dn}:\sum_{l=1}^n k_l=0,\sum_{l=j}^nk_l\in\bar V_m^+, j=2\ldots,n\}$.

\begin{lemma}
\label{3.2lem}
In a QFT with indefinite metric which fulfills the strong spectral condition, $\phi(\underline{\cal S}^+)\Omega$ is orthogonal (w.r.t. $(.,.)$) to $\Omega$ and is dense in the orthogonal complement of $\Omega$.
\end{lemma}
\noindent {\bf Proof.} We consider the following equation for $g\in\underline{\cal S}^+$ with $\underline{f}$ and $h$ as above:
\begin{eqnarray}
\label{3.3eqa}
0&=&  \hat h(0)\left\langle \Omega,\phi(\underline{f})\Omega\right\rangle
= \int_{\R^d}\left\langle \Omega,\phi(\underline{\alpha}_{\{1,a\}}(\underline{f}))\Omega\right\rangle h(a) da\nonumber \\
&=& \left\langle \Omega,\phi(\underline{g})\Omega\right\rangle=\left( \Omega,\eta\phi(\underline{g})\Omega\right).
\end{eqnarray}
Thus, $\eta\phi(\underline{g})\Omega$ is perpendicular to $\Omega$ w.r.t. $(.,.)$. Since $\eta\Omega=\Omega$ and $\eta$ is self adjoint, the same applies to $ \eta^2\phi(\underline{g})\Omega=\phi(\underline{g})\Omega$.

 It remains to show that vectors of the form $\phi(\underline{g})\Omega$ are dense
  in the orthogonal complement (w.r.t. $(.,.)$) of $\Omega$ in ${\cal H}$.
   By Lemma \ref{3.1lem}, the span of states of the form $(1-{\sf E}_0)\phi(\underline{f})\Omega$
    with $\underline{f}=(0,0,\ldots,f_r,0\ldots)$, with $\mbox{supp } \hat f_r$ compact,
	 is dense in the orthogonal complement of $\Omega$. We want to show that for such states
\begin{equation}
\label{3.4eqa}
(1-{\sf E}_0)\phi(\underline{f})\Omega=\phi(\underline{g})\Omega
\end{equation}
for some $\underline{g}=\underline{g}(\underline{f},h)\in\underline{\cal S}^+$, $h$ as above.
 To prove this, it is sufficient to show that the expectation values
  (w.r.t. $\langle.,.\rangle$) of both sides coincide for a set of vectors
   which span a dense set in ${\cal H}$, namely
    $\phi(\underline{q})\Omega$ with $\underline{q}=(0,0,\ldots,q_j,0,\ldots)$.
	 Taking this expectation value for the left hand side of (\ref{3.4eqa}) we get
$$
\left\langle\phi(\underline{q})\Omega,(1-{\sf E}_0)\phi(\underline{f})\Omega\right\rangle = \left\langle \Omega,\phi(\underline{q}^*\otimes\underline{f})\Omega\right\rangle-\left\langle\Omega,\phi(\underline{q}^*)\Omega\right\rangle\left\langle\Omega,\phi(\underline{f})\Omega\right\rangle
$$
where we used that $\eta\Omega=\Omega$. Expanding the right hand side into truncated Wightman functions according to Equation (\ref{3.2eqa}) and Fourier transforming we get for $n=r+j$
\begin{eqnarray*}
\ldots &=&
(2\pi)^{-dn}\int_{\R^{dn}}\sum_{\lambda\in{\cal P}(1,\ldots,n)}'\prod_{l=1}^{|\lambda|} \hat W^T_{|\lambda_l|}(k_{\lambda_l^1},\ldots,k_{\lambda_l^{|\lambda_l|}})\\
&\times&\hat q^*(k_1,\ldots,k_j)\hat f(k_{j+1},\ldots,k_n) dk_1\cdots dk_n
\end{eqnarray*}
where the reduced sum $\sum'$ runs over all partitions $\lambda$ s.t. there is at least one set $\lambda_l\in\lambda$ with $\lambda_l\cap (1,\ldots,j)\not=\emptyset$ and $\lambda_l\cap (j+1,\ldots,n)\not = \emptyset$.

 For $(k_1,\ldots,k_n)\in\mbox{supp }\prod_{l=1}^{|\lambda|} \hat W^T(\lambda_l)$ with $\lambda$ in the range of the sum, we obtain that $k=k_{j+1}+\ldots+k_n\in\bar V_{m}^+$ by the strong spectral condition fulfilled by 
 $\hat W_{|\lambda|}^T$ (in fact, $k$ is a sum of zero vectors and vectors from $\bar V_{m}^+$. Since $\lambda$ is in the range of the reduced sum, there must be at least one vector from $\bar V_{m}^+$ due to the support properties of $\hat W^T(\lambda_l)$).
  Thus, we can replace $\hat f_r(k_{j+1},\ldots,k_n)$ by $\hat h(k)\hat f_r(k_{j+1},\ldots,k_n)$ which does not change $\hat f_r$ on the support of $\prod_{l=1}^{|\lambda|}\hat W^T(\lambda_l)$. Furthermore, we can then drop the restriction of the sum and sum up over all partitions, since the terms which do not belong to the restricted sum give zero contribution if evaluated on $\hat q^*(k_1,\ldots,k_j)\hat h(k) \hat f(k_{j+1},\ldots,k_n)$ (on the support of such terms we have $k=0$).

If we now Fourier transform back, we get the $\langle.,.\rangle$-inner product of $\phi(\underline{q})\Omega$ with right hand side of Equation (\ref{3.4eqa}).
\kasten

\noindent {\bf Proof of Theorem \ref{3.3theo}:}
(i) We have to make sure that for $C\in{\cal B}({\cal H})$
$$
\left\langle \Psi_1,C\phi(\underline{f})\Psi_2\right\rangle=\left\langle\phi(\underline{f}^*)\Psi_1,C\Psi_2\right \rangle\forall \underline{f}\in\underline{\cal S},\Psi_1,\Psi_2\in{\cal D}
$$
implies $C=c1_{\cal H}$. Applying the above equation to
$\underline{g}^*$ where $\underline{g}\in\underline{\cal S}^+$
 and $\Psi_1=\Psi_2=\Omega$ we get that
$$
\left\langle \phi(\underline{g})\Omega,C\Omega\right\rangle=\left\langle\Omega,C\phi(\underline{g}^*)\Omega\right\rangle=0
$$
since $\underline{g}^*=\int_{\R^d}\underline{\alpha}_{\{1,a\}}(\underline{f}^*)\bar h(a)da\in I_{\rm sp}$ by the support properties of $\hat h$.

Since $\phi(\underline{\cal S}^+)\Omega$ is dense in the orthogonal complement of $\Omega$, we have thus obtained $C\Omega=c\Omega$. For $\Psi_1,\Psi_2\in{\cal D}, \Psi_2=\phi(\underline{f})\Omega$ for some $\underline{f}\in\underline{\cal S}$ we obtain
$$
\left\langle \Psi_1,C\Psi_2\right\rangle=\left\langle\phi(\underline{f}^*)\Psi_1,C\Omega\right\rangle=c\left\langle \Psi_1,\Psi_2\right\rangle
$$
and thus $C=c1_{\cal H}$.

(ii) Suppose $\tilde \Omega$ is translation invariant.
 We note that in Equation (\ref{3.3eqa}) we can replace the vector $\Omega$
  in the first argument of $\langle.,.\rangle$ and $(.,.)$
   respectively by $\tilde\Omega$ without changing the rest of the calculation.
    Thus, $\tilde\Omega$ is perpendicular (w.r.t. $(.,.)$) to
	 $\eta \phi(\underline{\cal S}^+)\Omega$. By Lemma \ref{3.2lem} $\phi(\underline{\cal S}^+)
	 \Omega$ is dense in the orthogonal complement of $\Omega$.
	  Since $\eta$ self adjoint and $\eta\Omega=\Omega$, this is also true for
	   $\eta  \phi(\underline{\cal S}^+)\Omega$. Thus $\tilde \Omega=c\Omega$.
\kasten

\section{Complex velocity transformations for the indefinite metric case}
In this section we prove the existence of a dense set of entire analytic vectors for the generator of the
velocity transformations (Lorentz boosts) in $x^0$ -- $x^1$ direction. This will then allow us to define the complex velocity transformations,
which are needed in the BW theorem to express the "modular operator", on this domain. In this way we avoid the use of functional calculus
which in general is not available on Krein spaces. The strategy is as follows: One first defines "Gaussian spectral cut-offs" for the Lorentz boosts  
on the Borchers algebra and thereby obtains an analytic action of the velocity transformations on a dense sub-*-algebra $\underline{\cal S}^{\rm anal.}$ of the Borchers algebra.
Then one uses Lemma \ref{3.1lem} to derive the existence of a dense set of analytic vectors ${\cal D}^{\rm anal.}\supseteq\phi(\underline{S}^{\rm anal.})\Omega$. 

We start with the definition of Gaussian mollifiers and their complex extension. Let, for $z\in\C$,
\begin{equation}
\label{4.1eqa}
c_\epsilon(z)=(\sqrt{2\pi\epsilon})^{-1}e^{-z^2/2\epsilon}~,~~\epsilon>0,
\end{equation}
such that $c_\epsilon(t)$ for $t\in\R$ is the centered Gaussian bell curve with
(variation) parameter $\epsilon>0$. We also note that $c_\epsilon*c_{\epsilon'}(t)
=c_{\epsilon+\epsilon'}(t)$ where $*$ means convolution in $t$, i.e. the $c_\epsilon$ form a convolution 
semigroup. As $\epsilon\downarrow 0$, $c_\epsilon(t)\to \delta(t)$ in the following sense:
\begin{lemma}
\label{4.1lem}
Let $F(t)$ be a continuous and exponentially bounded function, i.e. $\exists$ $C,M>0$ s.t. $|F(t)|\leq Ce^{M|t|}$ $\forall t\in \R$.
Then $\int_{\R}F(t)c_\epsilon(t)dt\to F(0)$ as $\epsilon\downarrow 0$. 
\end{lemma} 
\noindent {\bf Proof.} If $F$ is continuous and has compact support, the statement is well-known. Using
a partition of unity we can represent $F=F_1+F_2$ as a sum of a function $F_1$ with compact support
and a function $F_2$ which is zero on $[-1,1]$, thus $F(0)=F_1(0)$. It remains to show that the integral 
$\int_{\R}F_2(t)c_\epsilon(t)dt$ vanishes as $\epsilon\downarrow 0$. This is true by the theorem of dominated convergence as
$F_2(t)c_\epsilon(t)$ converge to zero pointwisely $\forall t\in \R$ and this function, for $\epsilon <1$, is dominated by the integrable function
$|F_2(t)|c_1(t)$.\kasten

Let $\alpha_t=\alpha_{\{\Lambda(t),0\}}$ with
\begin{equation}
\label{4.2eqa}
\Lambda(t)=\Lambda_{0,1}(t)=\left( 
\begin{array}{ccccc}
\cosh t & \sinh t & 0& \cdots &0\\
\sinh t & \cosh t &0 &\cdots &0 \\
0&0 & 1& \ddots&  \vdots\\
\vdots & &\ddots & \ddots  & 0\\
0& \cdots & &0&1
\end{array}\right)
\end{equation}
the velocity transformation in $x^0$ -- $x^1$ direction. By $|\Lambda(t)|$ we denote the maximum norm of the matrix
$\Lambda(t)$ and we note that this is an exponentially bounded function.
 
For $f\in{\cal S}_1$ and $\epsilon>0$ we set
\begin{equation}
\label{4.3eqa}
f_\epsilon=\int_{\R}\alpha_t(f)c_\epsilon(t)\, dt
\end{equation}
and we obtain the following lemma:
\begin{lemma}
\label{4.2lem}
Let $f\in S_1$. Then

\noindent (i) $f_\epsilon\in{\cal S}_1$ $\forall \epsilon>0$;

\noindent (ii) $f_\epsilon \to f$ in ${\cal S}_1$ as $\epsilon \downarrow 0$;
\end{lemma} 
\noindent {\bf Proof.} (i) Recall that ${\cal S}_1$ is the projective limit $\cap_{L,N\geq0}{\cal S}_{L,N}$
of $L$-time continuously differentiable functions on $\R^d$ with finite Schwartz norm $\|.\|_{L,N}$. It is thus sufficient
to prove that $f_\epsilon$ for all $\epsilon>0$ and $L,N\geq0$ has finite $\|.\|_{L,N}$ norm:
\begin{eqnarray}
\label{4.4eqa}
\|f_\epsilon\|_{L,N}&\leq&\int_{\R}\|\alpha_t(f)\|_{L,N}\, c_\epsilon(t)\, dt\nonumber\\
&\leq&\int_{\R} (1+|\Lambda(t)|^2)^{(N+L)/2}\,c_\epsilon(t)\, dt ~\|f\|_{L,N}
\end{eqnarray}
and we note that the integral on the right hand side is finite due to the Gaussian decay of $c_\epsilon(t)$ and
the only exponential increase of $|\Lambda(t)|$. Here we used the estimate
\begin{eqnarray}
\label{4.5eqa}
\|\alpha_t(f)\|_{L,N}&\leq&\sup_{x\in\R^d,|\beta|<L}\left| (1+|\Lambda(-t)x|^2)^{N/2}|\Lambda(t)|^{|\beta|}\partial^{\beta}f(x)\right|\nonumber\\
&\leq& (1+|\Lambda(t)|^2)^{(L+N)/2}\, \|f\|_{L,N}.
\end{eqnarray} 
(ii) As $\int_{\R} c_\epsilon(t)dt=1$,
$$
\left\|f_\epsilon-f\right\|_{L,N}\leq \int_{\R}\|\alpha_t(f)-f\|_{L,N}\, c_\epsilon(t)dt.
$$
Note that by (\ref{4.5eqa}) $F(t)=\| \alpha_t(f)-f\|_{L,N}$ is exponentially bounded in $t$. $F(t)$ is also continuous. Thus, 
by Lemma \ref{4.1lem}, we get for arbitrary $L,N\in\N_0$ that the right hand side of the above 
estimate converges to $F(0)=0$ for $\epsilon\downarrow 0$. 
As the family of Schwartz norms $\{\|.\|_{L,N}\}_{L,N\in\N_0}$ generates the Schwartz topology, the statement follows. \kasten

\begin{remark}
\label{4.1rem}
{\rm If we do not deal with a scalar theory, Lemma \ref{4.2lem} and all other constructions of this work
 can be generalized to representations $\alpha_{\{\Lambda,0\}}$ where
$\Lambda$ acts on the spin components via a finite-dimensional linear representation $\tau$ s.t. the matrix elements of $\tau(\Lambda(t))$
are exponentially bounded. This follows from $\tau(\Lambda(t))$ being additive and continuous in $t\in\R$. \kasten}
\end{remark}
Clearly, (\ref{4.3eqa}) can be interpreted as a Gaussian cut-off $\hat c_\epsilon(q)=e^{-\epsilon |q|^2}$ of
the spectrum of the generator of the velocity transformations on ${\cal S}_1\subseteq L^2(\R^d,\C)$. Lemma \ref{4.2lem} then proves that
such cut-offs (in contrast to a simple $L^2$-projection to a compact spectral set) take ${\cal S}_1$ to itself and can also be removed
without leaving ${\cal S}_1$. We next show that such cut-offs give the required analyticity on ${\cal S}_1$:

\begin{proposition}
\label{4.1prop}
Let $f\in{\cal S}_1$ and $\epsilon >0$. The function $\alpha_t(f_\epsilon)$ from $\R$ to ${\cal S}_1$ has an entire
analytic continuation $\alpha_z(f_\epsilon)$ from $\C$ to ${\cal S}_1$.
\end{proposition}
\noindent {\bf Proof.} We first note that for $f\in{\cal S}_1$, $\epsilon>0$, $\alpha_t(f_\epsilon)=\int_{\R}\alpha_s(f)\, c_\epsilon(s-t)\, ds$ and thus
the natural extention of $\alpha_t$, $t\in\R$, to complex parameters is given by
\begin{equation}
\label{4.6eqa}
\alpha_z(f_\epsilon)=\int_{\R} \alpha_s(f)\, c_\epsilon(s-z)\, ds,~~z\in\C.
\end{equation}
Since $c_\epsilon(z-s)=e^{-(z^2-2sz)/2\epsilon}c_\epsilon(s)$ and $e^{-(z^2-2sz)/2\epsilon}$ for $z\in\C$ fixed 
is exponentially bounded in $s$, one can show that $\alpha_z(f_\epsilon)\in{\cal S}_1$ $\forall z\in\C$ in analogy
to the proof of Lemma \ref{4.2lem} (i). 

Next we have to prove that $\alpha_z(f_\epsilon)$ is entire analytic in $z$, hence $\alpha_z(f_\epsilon)=\sum_{l=0}^\infty z^l f_l$ for $z\in\C$. 
Complex differentiability of $\alpha_z(f_\epsilon)$ then holds automatically and we get  $f_l=(1/l!)d^l\alpha_z(f_\epsilon)/dz^l|_{z=0}$.
For notational convenience we prove this statement for  $\tilde\alpha_z(f_\epsilon)=e^{z^2/2\epsilon}\alpha_z(f_\epsilon)$ which is equivalent. Note that
by an argument as in the proof of Lemma \ref{4.2lem} (i) $\int_{\R}\alpha_s(f)c_\epsilon (s) s^l ds\in{\cal S}_1$. What we have to show is thus that
for $L,N\in\N_0$
\begin{eqnarray}
\label{4.7eqa}
&& \left\|\tilde \alpha_z(f_\epsilon)-\sum_{l=0}^n{(z/\epsilon)^l\over l!}\int_{\R}\alpha_s(f)c_\epsilon (s)s^l ds\right\|_{L,N}\nonumber\\
&\leq&\sum_{l=n+1}^\infty {(|z|/\epsilon)^l\over l!} \int_{\R}\left\|\alpha_s(f)\right\|_{L,N}\, c_\epsilon (s) |s|^l ds\to 0 \mbox{ as } n\to\infty.
\end{eqnarray}
 We also note that by an calculation as (\ref{4.5eqa}) for $l$ even
 $$
 \int_{\R}\left\| \alpha_s(f)\right\|_{L,N}c_\epsilon (s) |s|^l ds\leq\int_{\R}c_\epsilon (s)|s|^l(1+|\Lambda(s)|^2)^{(N+L)/2}ds\, \|f\|_{L,N}
 $$
 Let $0<\epsilon<\epsilon'$. Then there exists a constant $C>0$ depending on $L,N$ and $\epsilon,\epsilon'$ such that the integral on the right hand
 side of the above estimate is smaller than $C\int_{\R}c_{\epsilon'}(s)|s|^lds$. Since 
 $$
 \sum_{l=0}^\infty {R^l\over l!}\int_{\R}c_{\epsilon'}(s)|s|^lds=\int_{\R}c_{\epsilon'}(s)e^{R|s|}ds<\infty
 $$ 
 for all $R>0$, (\ref{4.7eqa}) follows. \kasten
 
 We now define ${\cal S}_1^{\rm anal.}$ to be the $\C$-linear span of $\{\alpha_z(f_\epsilon):f\in{\cal S}_1,\epsilon>0,z\in\C\}$. Some properties of this space are listed in the following lemma:
 \begin{lemma}
 \label{4.3lem}
 (i) ${\cal S}_1^{\rm anal.}$ is a vector space which is closed under taking the complex conjugate;
 
 \noindent (ii) ${\cal S}_1^{\rm anal.}\subseteq {\cal S}_1$ is dense in ${\cal S}_1$;
 
 \noindent (iii) The mapping $\alpha_z:{\cal S}_1^{\rm anal.}\to {\cal S}_1^{\rm anal.}$ is well-defined for $z\in\C$.
 
 \noindent (iv)  $\C\ni z\to \alpha_z(f)\in{\cal S}_1^{\rm anal.}$, $f\in{\cal S}_1^{\rm anal.}$, is additive in $z$,
  i.e. $\alpha_{z_1}(\alpha_{z_2}(f))=\alpha_{z_1+z_2}(f)$ $\forall z_1,z_2\in\C$, and
  $\alpha_z(f)^*=\alpha_{\bar z}(f^*)$.
 
 \noindent (v) The mapping $\C\ni z\to \alpha_z(f)\in{\cal S}_1^{\rm anal.}$ for $f\in{\cal S}_1^{\rm anal.}$ is
 entire analytic (here we admit the coefficients for the expansion of $\alpha_z(f)$, $f\in{\cal S}^{\rm anal.}_1$, in $z$ to be in ${\cal S}_1$, not necessarily in ${\cal S}_1^{\rm anal.}$) in $z$.
 \end{lemma}
 \noindent {\bf Proof.} (i) Closedness under taking the complex conjugation is obvious, cf. (\ref{4.3eqa}).
 
 (ii) This follows from Lemma \ref{4.2lem} (i) and (ii).
 
 (iii) We first have to show that for $g\in{\cal S}_1^{\rm anal.}$, $g=\alpha_{z_1}(f_{\epsilon_1})=\alpha_{z_2}(h_{\epsilon_2})$, $f,h\in{\cal S}_1$, $\epsilon_1,\epsilon_2>0$, $z_1,z_2\in\C$,
 $\alpha_z(g)=\alpha_z(\alpha_{z_1}(f_{\epsilon_1}))=\alpha_{z+z_1}(f_{\epsilon_1})=\alpha_{z+z_2}(h_{\epsilon_2})=\alpha_z(\alpha_{z_2}(h_{\epsilon_2}))$ is not ambiguous. Note that this holds for real $z$. By entire analyticity in $z$, cf. Prop. \ref{4.1prop},
 $\alpha_{z}(\alpha_{z_2}(f_{\epsilon_1}))=\alpha_z(\alpha_{z_2}(h_{\epsilon_2}))$ then extends to all $z\in\C$.  
 
 (iv) That $\alpha_z$ is additive in $z$ and $\alpha_z(f)^*=\alpha_{\bar z}(f^*)$ follows from equation (\ref{4.6eqa}). 
 
 (v) is an immediate consequence of (iii) and Proposition \ref{4.1prop}.\kasten 

Let $\underline{\cal S}^{\rm anal.}$ be the free, unital, *-algebra generated by ${\cal S}_1^{\rm anal.}$, i.e.
\begin{equation}
\label{4.8eqa}
\underline{\cal S}^{\rm anal.}=\bigoplus_{n=1}^\infty {\cal S}_n^{{\rm anal.}}, ~~{\cal S}_n^{\rm anal.}={\cal S}_1^{{\rm anal.} \otimes n},~n\geq 1,~{\cal S}_0^{{\rm anal.}}=\C\, .
\end{equation}
By Lemma \ref{4.3lem} (i) this is well defined. Then $\underline{\cal S}_1^{\rm anal.}$ is a dense, unital sub-*-algebra of $\underline{\cal S}$, cf. 
Lemma \ref{4.3lem} (ii).

\begin{proposition}
\label{4.2prop}
(i) $\underline{\alpha}_{\, z}:\underline{\cal S}^{\rm anal.}\to\underline{\cal S}^{\rm anal.}$ is a well defined unital
algebra automorphism for $z\in\C$. It behaves under taking the involution as $\underline{\alpha}_{\, z}(\underline{f})^*=
\underline{\alpha}_{\, \bar z}(\underline{f}^*)$ and is entire analytic and additive in $z\in\C$.

\noindent (ii) Let $\underline{W}\in\underline{\cal S}'$ be Poincar\'e invariant  (cf. A2)) and $\underline{\cal S}^{\rm anal.}_0=\{ \underline{f}\in\underline{S}^{\rm anal.}:\underline{W}(\underline{g}\otimes\underline{f})=0\,\forall \underline{g}\in\underline{\cal S}\}$.
Then $\underline{\alpha}_{\, z}(\underline{\cal S}^{\rm anal.}_0)\subseteq\underline{\cal S}_0^{\rm anal.}$.
\end{proposition}
\noindent {\bf Proof.} (i) It is sufficient to prove this for the restriction of $\underline{\alpha}_{\, z}$ to ${\cal S}^{\rm anal.}_n$.
By Lemma \ref{4.3lem}, the statement holds for elements $\alpha_{z_1}(f_1)\otimes\cdots\otimes\alpha_{z_n}(f_n)$, $f_l\in{\cal S}_1^{\rm anal.}$, $z_l\in\C$, $l=1,\ldots,n$
in each of the components $\alpha_{z_l}(f_l)$. The claim now follows from going to the diagonal $z=z_1=\ldots=z_n$.

(ii) Let $\underline{f}\in\underline{\cal S}^{\rm anal.}_0$. Then $\underline{W}(\underline{g}\otimes\alpha_z(\underline{f}))=0$ for all real $z$. By (i) and $\underline{W}\in\underline{\cal S}'$, $\underline{W}(\underline{g}\otimes\alpha_z(\underline{f}))$
is an analytic function in $z$ and thus vanishes identically.
\kasten

 We now want to use Lemma \ref{3.1lem} in order to show that the results of Proposition \ref{4.2prop} carry over to the Krein space representation on ${\cal H}$. We first give some definitions:
 \begin{definition}
 \label{4.1def}
 {\rm
 Let $({\cal H},\eta)$ be a Krein space with fundamental decomposition $\eta$ and let ${\sf U}:\R\to{\sf O}_\eta({\cal D})$ be a 
 one parameter group of $\eta$-unitary operators ${\sf U}(t):{\cal D}\to{\cal D}$. 
 
 \noindent (i) Let
 $${\cal D}({\sf A})=\left\{ \Psi\in{\cal D}:\lim_{t\to0, t\not=0}{{\sf U}(t)-1\over it}\,\Psi\mbox{\rm ~converges strongly in~}{\cal H}\right\}.$$
 Then ${\sf A}=({\sf A},{\cal D}({\sf A}))$, ${\sf A}\Psi=\lim_{t\to0, t\not=0}(it)^{-1}({\sf U}(t)-1)\Psi$, $\Psi\in{\cal D}({\sf A})$, is called the generator of $\{{\sf U}(t)\}_{t\in\R}$.

 \noindent (ii) A vector $\Psi\in{\cal D}({\sf A})$ is called analytic for ${\sf A}$ if  $\forall n\in\N$ ${\sf A}^n\Psi\in{\cal D}({\sf A})$ and
 $\exists R>0$ such that $\forall z\in\C$, $|z|<R$, $\sum_{n=0}^\infty (z^n/n!){\sf A}^n\Psi$ converges strongly in ${\cal H}$. If this holds for all $R>0$,
 $\Psi$ is called entire analytic.
 
 \noindent (iii) By ${\cal D}^{\rm anal.}={\cal D}^{\rm anal.}({\sf A})$ we denote the set of all entire analytic vectors of ${\sf A}$.
 }\kasten
 \end{definition}
 By standard methods one shows that $\sf A$ is $\eta$-symmetric on ${\cal D}({\sf A})$.
 By these very definitions we get ${\sf U}(z)=\sum_{n=0}^\infty((iz)^n/n!){\sf A}^n:{\cal D}^{\rm anal.}\to{\cal H}$ is well-defined and additive for $z\in\C$ and ${\sf U}(z)\Psi$, $\Psi\in{\cal D}^{\rm anal.}$, is entire analytic
 in $z\in\C$ in the strong topology on ${\cal H}$.
 
 We now get as the main theorem of this section:
 \begin{theorem}
 \label{4.1theo}
 Let $\underline{W}\in{\cal S}'$ fulfill the Axioms \ref{2.1ax} with Sobolev dominated Hilbert seminorm $\underline{p}$ and let $(({\cal H},\langle.,.\rangle,\Omega),\eta,{\sf U},\phi)$ be
 the associated QFT with indefinite metric, cf. Theorem \ref{2.1theo}. Let, furthermore,  ${\sf U}(t)={\sf U}(\{\Lambda(t),0\})$, $t\in\R$ and ${\sf A}$ be the generator of the one parameter $\eta$-unitary group $\{{\sf U}(t)\}_{t\in\R}$. 
 
 Then ${\sf A}$ has a dense domain ${\cal D}^{\rm anal.}$ of entire analytic vectors.
 In particular, $\{{\sf U}(t)\}_{t\in\R}$ on ${\cal D}^{\rm anal.}$ has a well-defined additive extension to ${\sf U}(z)$, $z\in\C$, given
 by ${\sf U}(z)=\sum_{n=0}^{\infty}((iz)^n/n!){\sf A}^n$ and ${\sf U}(z)^{[*]}={\sf U}(-\bar z)$ holds on this domain.  

  \end{theorem} 
 \noindent {\bf Proof.} As $\underline{\cal S}^{\rm anal.}$ is dense in $\underline{\cal S}$, it follows from Lemma
 \ref{3.1lem} and the fact that $\phi(\underline{\cal S})\Omega$ is dense in ${\cal H}$ that also
 $\phi(\underline{\cal S}^{\rm anal.})\Omega$ is dense in ${\cal H}$.
 
  It therefore suffices to show that
 $\phi(\underline{\cal S}^{\rm anal.})\Omega\subseteq{\cal D}^{\rm anal.}$. Let $\Psi\in\phi(\underline{\cal S}^{\rm anal.})\Omega$,
 $\Psi=\phi(\underline{f})\Omega$,
 $\underline{f}\in\underline{\cal S}^{\rm anal.}$. 
 By Proposition \ref{4.2prop} (ii),
 ${\sf U}(z)\Psi=\phi(\underline{\alpha}_{\, z}(\underline{f}))\Omega$ is well-defined.  Furthermore, by Prop. \ref{4.2prop} $\exists$ $\underline{f}_n\in\underline{\cal S}$ such that
 $\underline{\alpha}_{\, z}(\underline{f})=\linebreak \sum_{l=0}^{\infty}((iz)^n/n!)\underline{f}_n$. Hence, $\lim_{t\to0,t\not=0}(it)^{-1}(\underline{\alpha}_{t}(\underline{ f})-\underline{f})=\underline{f}_1$ in $\underline{\cal S}$.
 By Lemma \ref{3.1lem} this implies
 $$
 \phi(\underline{f}_1)\Omega=\phi\left(\lim_{t\to0,t\not=0}{\alpha_{t}(\underline{f})-\underline{f}\over it}\right)\Omega=\lim_{t\to0,t\not=0}{{\sf U}(t)-1\over it}\, \Psi\,.
 $$
 Thus $\Psi\in {\cal D}({\sf A})$ and ${\sf A}\Psi=\phi(\underline{f}_1)\Omega$. Repeating this argument $n$ times we get that ${\sf A}^{n-1}\Psi\in{\cal D}({\sf A})$ and
 ${\sf A}^{n}\Psi=\phi(\underline{f}_n)\Omega$. Again by Lemma \ref{3.1lem} we get that the following chain of equation holds: 
 \begin{eqnarray*}
 {\sf U}(z)\Psi=\phi\left(\underline{\alpha}_{\, z}(\underline{f})\right)\Omega
 &=&\phi\left(\lim_{N\to\infty}\sum_{n=0}^N{(iz)^n\over n!}\underline{f}_n\right)\Omega\\
 &=&\lim_{N\to\infty}\phi\left(\sum_{n=0}^N{(iz)^n\over n!}\underline{f}_n\right)\Omega\\
 &=&\lim_{N\to\infty}\sum_{n=0}^{N}{(iz)^n\over n!}{\sf A}^n\Psi\, , 
 \end{eqnarray*}
 where the convergence is strong convergence in ${\cal H}$ and $z\in\C$ is arbitrary. This, by definition of ${\cal D}^{\rm anal.}$, just means that $\Psi\in{\cal D}^{\rm anal.}$. \kasten

Let us briefly discuss the relation of Therorem \ref{4.1theo} with functional calculus on Krein spaces: Setting ${\cal C}=\{ \xi :\C\to\C, \xi(z)=\sum_{l=1}^\infty a_lz^l $  
$a_l\in\C$ s.t. $\exists R>0,C>0$, $|a_l|<CR^l/l!\forall l\in\N\}$, we get by the same methods as in the proof of Proposition \ref{4.1prop} and Theorem \ref{4.1theo} that
$(F({\sf A}),\phi(\underline{\cal S}^{\rm anal.})\Omega)$ is well-defined for all $F$ in the algebra ${\cal C}$. This defines a restricted functional calculus for ${\sf A}$. But this algebra of entire analytic functions of course does not contain
any characteristic sets ("spectral projections").

If one would try to use the approach of this paper to define spectral projections, one essentially has to continuously extend the Wightman functional $\underline{W}$ to a bigger algebra generated by elements
\begin{equation}
\label{4.9eqa}
\int_{\R}\alpha_{ t}(f) c_{[a,b]}(t)\,dt~,~~f\in{\cal S}_1
\end{equation} 
where $c_{[a,b]}$ is the inverse Fourier transform of the characteristic function of the set $[a,b]$. While it is easy to show that this defines a function on $\R^d\setminus\{x\in\R^d:|x^0|=|x^1|\}$, the estimates
of the proof of Lemma \ref{4.2lem} obviously fail as $c_{[a,b]}(t)$ falls to zero no faster than $1/t$ as $t\to\infty$. One thus would have to restrict to Wightman functionals $\underline{W}$ with
Wightman functions being rather functions of sufficiently fast decay than just distributions. This however is not a physical assumption, as e.g. the two-point function of a scalar field in the physical space-time $d=4$
has singularities on the light cone  $\sim 1/x^2$ and thus is not a locally integrable function and due to Lorentz invariance is constant along the orbits of the Lorentz group. This forced us in (\ref{4.3eqa}) to work with the Gaussian
cut-offs giving a less sharp localization of the spectrum. Of course, spectral projections in some cases could be defined using different methods. 

\begin{remark}
\label{4.2rem}
{\rm On an  informal level, spectral calculus can also be defined using Gaussian localization of the spectrum $\hat c_{a,\epsilon}({\sf A})=\sqrt{\epsilon\over
 \pi}e^{-\epsilon|a-{\sf A}|^2}$, $a\in\R,\epsilon>0$. Informally, $\lim_{\epsilon\uparrow\infty}\hat c_{a,\epsilon}({\sf A})=d{\sf E}_{(-\infty,a]}/da$ where ${\sf E}_{(-\infty,a]}=\hat c_{(-\infty,a]}({\sf A})$
is the spectral projection of ${\sf A}$ on $(-\infty,a]$. Thus one gets the following heuristic formula 
\begin{equation}
\label{4.10eqa}
F({\sf A})="{\int_{\R}F(a)\, d{\sf E}_{(-\infty,a]}}"=\lim_{\epsilon\uparrow\infty}\int_{\R}F(a)\, \hat c_{a,\epsilon}({\sf A})\,da
\end{equation}
and one could study the existence of this limit depending on the choice of $F$ and suitable domains in order to extend the choice of admissible $F$ beyond ${\cal C}$. \kasten
}
\end{remark}

\section{On the BW theorem for QFTs with indefinite metric} 
In this section we combine the existence of complex velocity transformations established in the previous section with a theorem of 
J. Bros, H. Epstein and U. Moschella which shows that Bisognano Wichmann analyticity
(in particular: the KMS-condition) does not depend on positivity. We first recall this result, which
 is the crucial step towards the BW theorem for QFT with indefinite metric. Here ${\cal W}_{R}=\{ x\in\R^d:x^1>|x^0|\}$, ${\cal W}_L=-{\cal W}_R$  and $\Lambda(e_1,\pi)$ is the rotation 
 around $e_1=(0,1,0,\ldots,0)\in\R^d$ by the angle $\pi$. We also note that $\underline{\alpha}_{\, t}$ takes the *-algebras $\underline{\cal S}({\cal W}_{R/L})$ into themselves. 
 We only formulate our statements for the right wedge -- the related statement for the left wedge can be derived analogously.

\begin{theorem}[\cite{BEM}]
\label{5.1theo}
Let $\underline{W}\in\underline{\cal S}'$ fulfill temperedness, locality and the spectral condition (cf. A1), A3) and A4) of Axioms \ref{2.1ax}). Then,

\noindent (i) The "dynamical systems" $(\underline{\cal S}({\cal W}_{R}),\underline{\alpha}_{\, t},\underline{W})$ fulfills the KMS condition with temperature $1/2\pi$, i.e. $\forall \underline{f},\underline{g}\in\underline{\cal S}({\cal W}_{R})$
$\exists$ a holomorphic function $F(z)$ on the strip $\R+ i (0,2\pi)$
which continuously extends to $\R+ i[0,2\pi]$ such that 
$$
F(t)=\underline{W}(\underline{f}\otimes\underline{\alpha}_{\, t}(\underline{g})),~~ \mbox{ and }~~ F(t+i2\pi)=\underline{W}(\underline{\alpha}_{\, t}(\underline{g})\otimes\underline{f})~,~~t\in \R.
$$ 

\noindent (ii) For $F$ as in (i), the Bisognano--Wichmann relation is fulfilled: 
$$
F(i\pi)=\underline{W}(\underline{f}\otimes\underline{\alpha}_{\{\Theta_{0,1},0\}}(\underline{g}^*))~.
$$
\end{theorem}
\begin{remark}
\label{5.1rem}
{\rm The proof for this theorem, based on analytic completion along the orbits of the group of complex Lorentz boosts, has been formulated for the case of de Sitter space-time, but the same proof also holds for Minkowski space-time, cf. \cite{BEM}. 
Here, up to different notation, (i) corresponds to Eq. (45) of \cite{BEM} and (ii) is the special case $\lambda=-1$ of equation (46)
of that reference where $\lambda=e^z$, $z$ being the argument of $F$.

It should also be remarked that the statement of \cite{BEM} has been formulated for the
case of test functions with compact support. Nevertheless, the proof of the crucial Lemma 2 of that
reference is based on the estimate (67) which guarantees in (69) of \cite{BEM} boundary values in ${\cal S}'$, see Theorem 2-10 
of \cite{SW}.
But in the Minkowski case, (67) follows automatically from the assumption of temperedness Theorem 2-10 of \cite{SW} (see also (17) of \cite{BEM}). Hence the statement of Theorem \ref{5.1theo} also holds for Schwartz test functions. }\kasten
\end{remark} 

If we re-write (ii) of Theorem \ref{5.1theo} for a Wightman functional $\underline{W}$ associated to some quantum field theory with indefinite metric, cf. Theorem \ref{2.1theo}, we find 
\begin{equation}
\label{5.1eqa}
\langle \phi(\underline{f}^*)\Omega,{\sf U}(i\pi)\phi(\underline{g})\Omega\rangle=\langle\phi(\underline{f}^*)\Omega,{\sf J}\phi(\underline{g}^*)\Omega\rangle,~~~{\sf J}={\sf U}(\{\Theta_{0,1},0\}),
\end{equation}
where $\underline{f},\underline{g}\in\underline{\cal S}({\cal W}_R)$. ${\sf J}$ is a $\eta$-antiunitary and antiunitary (cf. Corollary \ref{2.1cor}) operator and ${\sf J}^2=1$. (\ref{5.1eqa}) is however only symbolic as it is not clear whether 
one can make sense out of the expression ${\sf U}(i\pi)\phi(\underline{g})\Omega$ for all $\underline{g}\in\underline{\cal S}({\cal W}_R)$. We therefore have to consider
a wedge algebra which is smaller than ${\cal P}_{R}=\phi(\underline{\cal S}({\cal W}_R))$
 and maps $\Omega$ to ${\cal D}^{\rm anal.}$.
  At the same time, such a smaller wedge algebra must not be too small as important properties, e.g. the Reeh-Schlieder property, have to be preserved
  and the weak commutant should coincide with the wedge algebra. Here we suggest the following solution:

Let $\underline{\cal S}^{\rm anal.}({\cal W}_{R/L})=\underline{\cal S}^{\rm anal.}\cap\underline{\cal S}({\cal W }_{R/L})$
 and ${\cal P}_{R/L}^{\rm anal.}=\phi(\underline{\cal S}^{\rm anal.}({\cal W}_{R/L}))$.
  By definition ${\cal P}_{R/L}^{\rm anal.}\subseteq {\sf O}_\eta({\cal D})$ and ${\cal P}^{\rm anal.}_{R/L}\Omega\subseteq {\cal D}^{\rm anal.}$.

The following properties of ${\cal P}^{\rm anal.}_{R/L}$ show that
these algebras can be seen as legitimate substitutes for the wedge algebras ${\cal P}_{R/L}$:

\begin{proposition}
\label{5.1prop} Let ${\cal P}^{\rm anal.}_{L/R}$ be defined as above. Then,

\noindent (i) $\Omega$ is standard (cyclic and separating) for ${\cal P}^{\rm anal.}_{R/L}$;

\noindent (ii) ${\cal P}^{\rm anal.}_{R/L}$ is dense in ${\cal P}_{R/L}$ w.r.t. the separating norm ${\cal P}_{R/L}\ni{\sf L}\to\|{\sf L}\Omega\|$;

\noindent (iii) The weak commutants of ${\cal P}_{R/L}$ and ${\cal P}^{\rm anal.}_{R/L}$ coincide.
\end{proposition}  
\noindent {\bf Proof.} (i) By Theorem \ref{3.2theo} $\Omega$ is cyclic and separating for ${\cal P}_{R/L}$. As ${\cal P}_{R/L}^{\rm anal.}\subseteq {\cal P}_{R/L}$, $\Omega$ 
is separating for ${\cal P}_{R/L}^{\rm anal.}$. It remains to show that $\Omega$ is also cyclic. Note that ${\cal S}_1({\cal W}_{L/R})$ is mapped into itself by the action of $\alpha_t$. 
Thus, as in Lemma \ref{4.2lem} (ii), one can show that ${\cal S}_1^{\rm anal.}({\cal W}_{R/L})={\cal S}^{\rm anal.}_1\cap{\cal S}_1({\cal W}_{R/L})$ is dense in ${\cal S}_1({\cal W}_{R/L})$.
Thus, $\underline{\cal S}^{\rm anal.}({\cal W}_{R/L})$ is dense in $\underline{\cal S}({\cal W}_{R/L})$. Consequently, by Lemma \ref{3.1lem}, any vector $\Psi=\phi(\underline{f})\Omega$, $\underline{f}\in\underline{\cal S}({\cal W}_{R/L})$,
can be approximated in the strong sense by a sequence of vectors $\Psi_n=\phi(\underline{f}_n)\Omega$, $\underline{f}_n\in\underline{\cal S}^{\rm anal.}({\cal W}_{R/L})$.

(ii) As we only have to show that any vector $\Psi={\sf L}\Omega$, ${\sf L}\in{\cal P}_{R/L}$,  can be approximated in the strong topology by vectors in $\{{\sf L}\Omega:{\sf L}\in{\cal P}^{\rm anal. }_{R/L}\}$, the same argument as in (i) can be used.

(iii) Obviously ${\cal P}'_{R/L}\subseteq{\cal P}^{{\rm anal.}'}_{R/L}$. We have to show the opposite inclusion. Let $C\in{\cal P}^{{\rm anal.}'}_{R/L}$. Then, by Lemma \ref{3.1lem}, we get for $\Psi_1,\Psi_2\in{\cal D}$, $\Psi_1=\phi(\underline{f})\Omega$, $\Psi_2=\phi(\underline{h})\Omega$, $\underline{f},\underline{h}\in\underline{\cal S}$ and ${\sf L}\in{\cal P}_{R/L}$, ${\sf L}=\phi(\underline{g})$, $\underline{g}\in \underline{\cal S}({\cal W}_{R/L})$
\begin{eqnarray*}
\langle\Psi_1,C{\sf L}\Psi_2\rangle &=&\langle\phi(\underline{f})\Omega,C\phi(\underline{g})\phi(\underline{h})\Omega\rangle\\
&=&\langle\phi(\underline{f})\Omega,C\phi(\underline{g}\otimes\underline{h})\Omega\rangle\\
&=&\lim_{n\to\infty}\langle\phi(\underline{f})\Omega,C\phi(\underline{g}_n\otimes\underline{h})\Omega\rangle\\
&=&\lim_{n\to\infty}\langle\Psi_1,C{\sf L}_n\Psi_2\rangle\\
&=&\lim_{n\to\infty}\langle{\sf L}_n^{[*]}\Psi_1,C\Psi_2\rangle=\langle{\sf L}^{[*]}\Psi_1,C\Psi_2\rangle
\end{eqnarray*} 
where in the last step one has to repeat the argument of the first four steps in reverse order to see that the equality follows from Lemma \ref{3.1lem}. Here we used that there exists a sequence $\underline{g}_n\in\underline{\cal S}^{\rm anal.}({\cal W}_{R/L})$ such that
$\underline{g}_n\to\underline{g}$ in $\underline{\cal S}$ as $n\to\infty$ (cf. the Proof of (i)) and ${\sf L}_n=\phi(\underline{g}_n)\in{\cal P}_{R/L}^{\rm anal.}$.
\kasten    

We define the domain ${\cal D}_{R}^{\rm anal.}={\cal P}_{R}^{\rm anal.}\Omega$. By Proposition \ref{5.1prop}, ${\cal D}^{\rm anal.}_{R}$ is dense in ${\cal H}$ and
\begin{equation}
\label{5.3eqa}
{\sf S}_{R}:{\cal D}^{\rm anal.}_{R}\to {\cal D}^{\rm anal.}_{R}\subseteq {\cal H},~~~{\sf S}_{R}:{\sf L}\Omega\mapsto{\sf L}^{[*]}\Omega, ~~{\sf L}\in{\cal P}_{R}^{\rm anal.}
\end{equation}
densely defines an anti-linear operator ${\sf S}_{R}$. Note that $\Omega$ is separating for ${\cal P}_{R}^{\rm anal.}$, hence (\ref{5.3eqa}) is not ambiguous. 
Also, $({\sf S}_{R}^{[*]},{\cal D}_{R}^{{\rm anal.}[*]})\supseteq({\sf S}_{L},{\cal D}^{\rm anal.}_L)$, with ${\sf S}_L$ defined in analogy to (\ref{5.3eqa}) for the right wedge replaced by the left wedge. 
This implies that ${\sf S}_{R}$ is closable.

We can now collect the pieces and formulate a version of the Bisonano--Wichmann theorem for quantum fields acting on Krein spaces:

\begin{theorem}
\label{5.2theo}
Let ${\cal P}_{R/L}^{\rm anal.}$ be the analytic right/left wedge algebras associated with some quantum field theory with indefinite metric and let
${\sf J}$ and ${\sf S}_{R}$ be defined as in Eq. (\ref{5.1eqa}) and (\ref{5.3eqa}). Then,

\noindent (i) $\left[{\cal P}_R^{\rm anal.},{\cal P}_L^{\rm anal.}\right]=0$ on ${\cal D}$;

\noindent (ii) ${\sf J}{\cal P}_R^{\rm anal.}{\sf J}={\cal P}_L^{\rm anal.}$ and ${\sf J}{\cal P}_L^{\rm anal.}{\sf J}={\cal P}_R^{\rm anal.}$;

\noindent (iii) ${\sf U}(t){\cal P}_{R/L}^{\rm anal.}{\sf U}(t)^{-1}={\cal P}_{R/L}^{\rm anal.}$;

\noindent (iv) ${\sf S}_R={\sf J}{\sf U}(i\pi)|_{{\cal D}_R^{\rm anal.}}$;

\noindent (v) $({\cal P}_R^{\rm anal.},{\sf U}(t), \Omega)$ fulfill the KMS condition (w.r.t. $\langle.,.\rangle$).
\end{theorem}
\noindent {\bf Proof.} (i) holds by locality.
(ii) This is a simple consequence of 
$$
\underline{\alpha}_{\{\Theta_{0,1},0\}}(\underline{\cal S}^{\rm anal.})=\underline{\cal S}^{\rm anal.}
~\mbox{ and }~
 \underline{\alpha}_{\{\Theta_{0,1},0\}}(\underline{\cal S}({\cal W}_{R/L}))=\underline{\cal S}({\cal W}_{L/R}).
 $$
 
 (iii) Similar as in (ii), this point follows from
 $$
 \underline{\alpha}_{\, t}(\underline{\cal S}^{\rm anal.})=\underline{\cal S}^{\rm anal.}
 ~\mbox{ and }~ 
 \underline{\alpha}_{\, t}(\underline{\cal S}({\cal W}_{R/L}))=\underline{\cal S}({\cal W}_{R/L}).
 $$

 (iv) We note that for $\underline{f},\underline{g}\in\underline{\cal S}^{\rm anal.}({\cal W}_R)$, $\tilde F(z)=\langle \phi(\underline{f}^*)\Omega,{\sf U}(z)\phi(\underline{g})\Omega\rangle$ defines
  an entire analytic function in $z$, cf. Theorem \ref{4.1theo}. This function $\tilde F(z)$ coincides for real $z$ with the continuation of the  analytic function $F(z)$ of Theorem \ref{5.1theo} (i).
  By the edge of the wedge theorem, $\tilde F(z)$ is thus
  an analytic continuation of $F(z)$ to all $\C$. We identify these two functions. For $z=i\pi$ we then get by Theorem \ref{5.1theo} (ii) that (\ref{5.1eqa}) holds rigorously. Thus,
  $$
  \langle \Psi_1, {\sf U}(i\pi){\sf L}\Omega\rangle=\langle\Psi_1,{\sf J}{\sf L}^{[*]}\Omega\rangle,~~\forall \Psi_1\in{\cal D}^{\rm anal.}_R,~{\sf L}\in{\cal P}_R^{\rm anal.}.
  $$
  $\Omega$ is cyclic for ${\cal P}_R^{\rm anal.}$, cf. Proposition \ref{5.1prop} (i), hence ${\cal D}_R^{\rm anal.}$ is dense in ${\cal H}$ and one gets
  from the above equation that ${\sf U}(i\pi)\Psi={\sf J}{\sf S}_R\Psi$ $\forall \Psi\in{\cal D}_R^{\rm anal.}$. The assertion now follows by multiplication of both sides with ${\sf J}$ using
  ${\sf J}^2=1$.
  
  (v) Like in (iv) one shows that the function $F(z)$ for suitable choice of $\underline{f},\underline{g}\in \underline{\cal S}^{\rm anal.}({\cal W}_R)$ coincides with $\langle\Omega,{\sf L}_1{\sf U}(z){\sf L}_2\Omega\rangle$, ${\sf L}_1,{\sf L}_2\in{\cal P}_R^{\rm anal.}$ and
  one then gets the KMS condition 
  $$
  \langle\Omega,{\sf L}_1{\sf U}(t+i2\pi){\sf L}_2\Omega\rangle=\langle\Omega,{\sf L}_2{\sf U}(-t){\sf L}_1\Omega\rangle
  $$ 
  from Theorem \ref{5.1theo} (i). \kasten 

 \

\noindent {\bf Acknowledgments.} Interesting and helpful discussions with S. Albeverio, G. Hofmann, F. Ll\'edo, G. Morchio, G. Piacitelli, F. Strocchi and J. Yngvason are gratefully acknowledged. 
I have to thank the referee for the very careful reading of the paper. This work has been made possible
through financial support by the D.F.G. project "Stochastic analysis and systems with infinitely many degrees of freedom".


\begin{thebibliography}{11}
\bibitem{AG} S. Albeverio, H. Gottschalk: Scattering theory for quantum fields with indefinite metric, Commun. Math. Phys. {\bf 261}, 491--513 (2001).
\bibitem{AGW1} S. Albeverio, H. Gottschalk, J.-L. Wu, { Convoluted generalized white noise, Schwinger functions and their continuation to Wightman functions}, Rev. Math Phys., Vol {\bf 8}, No. {\bf 6}, 763--817, (1996).
\bibitem{AGW2} S. Albeverio, H. Gottschalk, J.-L. Wu, { Models of local relativistic quantum fields with indefinite metric (in all dimensions)}, Commun. Math. Phys. {\bf 184}, 509--531, (1997).
\bibitem{An} T. Ya. Azizov, I. S. Iokhvidov, Linear Operators in spaces with an indefinite metric, Wiley-Interscience, Chichester / N. Y., 1989.
\bibitem{BWo} H. Baumg\"artel, M. Wollenberg, Causal nets of operator algebras. Mathematical aspects of algenraic quantum field theory, Akademie-Verlag, Berlin, 1992.
\bibitem{BW} J. J. Bisognano, E. H. Wichmann, On the duality conditions for a Hermitian scalar field, Journ. Math. Phys. {\bf 16} No. {\bf 4}, 985-1007 (1975).
\bibitem{Bon} J. Bogn\'ar, Indefinite inner product spaces, Springer, Heidelberg / N. Y. 1974.
\bibitem{Bo1} H. J. Borchers, On the structure of the algebra of field operators, Il Nuovo Cimento {\bf 24}, 214--236 (1962).
\bibitem{Bo2} H. J. Borchers, Algebraic aspects of Wightman field theory, pp. 31-80 in Statistical mechanics and field theory, Ed. R. N. Sen and C. Weil, Halsted Press, New York/Israel Universities Press Jerusalem/London, 1972.
\bibitem{Bo3} H. J. Borchers, On revolutionizing quantum field theory with Tomita's modular theory, Journ. Math. Phys. {\bf 41}, No. {\bf 6}, 3604-3673 (2000).
\bibitem{BR} O. Bratelli, D. W. Robinson, Operator algebras and quantum statistical mechanics I, 2nd edition, Springer Verlag N.Y. / Berlin / Heidelberg, 1987.
\bibitem{BEM} J. Bros, H. Epstein, H. Moschella, Analyticity properties and thermal effects for general quantum field 
theory on de Sitter space-time, Commun. Math. Phys. {\bf 196}, 535--570 (1998).
\bibitem{Bu1} D. Buchholz, The physical state space of quantum electrodynamics, Commun. Math. Phys. {\bf 85}, 49-71 (1982). 
\bibitem{Bu} D. Buchholz, S. Doplicher, R. Longo, J. E. Roberts, A new look at Goldstone's theorem, Rev. Math. Phys., Secial Issue dedic. R. Haag, p. 49-83 (1992). 
\bibitem{CG} T. Constantinescu, A. Gheondea, Representations of Hermitian kernels by means of Krein spaces II: Invariant kernels, Commun. Math. Phys. {\bf 216},  409--430 (2001).
\bibitem{Do} S. Doplicher, An algebraic spectral condition, Commun. Math. Phys. {\bf 1}, 1--5 (1965).
\bibitem{GV} I. M. Gelfand, N. Ya. Vilenkin, Generalized Functions, Vol. I-IV, Academic Press 1964.
\bibitem{Ha} R. Haag, Local quantum physics, fields particles algebras, Springer, Berlin 1992.
\bibitem{HL} A. Dijksma, I. Gohberg, M. A. Kaashoek, R. Menniken (Eds.), Contributions to Operator theory in spaces with indefinite metric, Adv. Appl. Operator theory, Vol. {\bf 106}, Birkh\"auser 1998. 
\bibitem{Ho1} G. Hofmann, {The Hilbert space structure condition for quantum field theories with indefinite metric and transformations with linear functionals}, Lett. Math. Phys. {\bf 42}, 281--295, (1997).
\bibitem{Ho2} G. Hofmann, { On GNS representations on inner product spaces: I. The structure of the representation space}, Commun. Math. Phys., {\bf 191},  299--323 (1998).
\bibitem{Ho3} G. Hofmann, On the characterization of Pseudo Krein and Pre-Krein spaces, preprint Leipzig, 1999.
\bibitem{HoX} G. Hofmann, K. Luig, On the construction of selfpolar and selfpolar Hilbertian norms on inner product spaces, preprint NTZ, Leipzig 2000.
\bibitem{In} A. Inoue, Tomita--Takesaki theory in algebras of unbounded operators, Lect. Notes Math. {\bf 1699}, Springer Verlag, Berlin 1999. 
\bibitem{KS} E. Kissin, V. Shulman, Representations on Krein spaces and derivations of $C^*$ algebras, Pitman Monographs Pure Appl. Math. {\sf 89}, Longman Harlow 1997.
\bibitem{MS} G. Morchio, F. Strocchi, { Infrared singularities, vacuum structure and pure phases in local quantum field theory}, Ann. Inst. H. Poincar\'e, Vol. {\bf 33},  251--282, (1980).
\bibitem{MS1} G. Morchio, F. Strocchi, A non-perturbative approach to the infared problem in QED: Construction of charged states. Nuclear Physics {\bf B 211}, 471--508 (1983).
\bibitem{MS2} G. Morchio, F. Strocchi, Representations of *-algebras in indefinite inner product spaces. in: Proc. Conf. "Quantum theory and stochastic analysis, new interplays" in Honor 60. Birthday S. Albeverio, Eds. Gesetzky/Holden/Jost/Paycha,
Can. Math. Soc. {\bf 29}, 491-503 (2000).
\bibitem{To} N. M. Nikolov, I. T. Todorov, Rationality of conformally invariant local correlation functions on compactified Minkowski space, Commun. Math. Phys. {\bf 218} 417--436 (2001).
\bibitem{Os} A. Ostendorf, Feynman rules for Wightman functions, Ann Inst. H. Poincar\'e {\bf 40}, 273--290 (1984).
\bibitem{SW} R. F. Streater, A. S. Wightman, PCT, spin \& statistics and all that\ldots, Benjamin, New York 1964.  
\bibitem{Ste1} O. Steinmann, Perturbation theory of Wightman functions, Commun. Math. Phys. {\bf 152}, 627--645 (1993).
\bibitem{Ste2} O. Steinmann, Perturbative quantum electrodynamics and axiomatic field theory, Springer Berlin/Heidelberg/N.Y., 2000.
\bibitem{Str1} F. Strocchi, {Local and covariant gauge quantum field theories. Cluster property, superselection rules and the infrared problem}, Phys. Rev. {\bf D 17}, 2010--2021 (1978).
\bibitem{Str3} F. Strocchi, Locality and covariance in QED and gravitation: General proof of Gupta--Bleuler type formulations, Mathematical Methods in Theoretical Physics, W. E. Brittin (ed.), Colorado Ass. Univ. Press, Boulder 1973.
\bibitem{Str2} F. Strocchi, { Selected topics on the general properties of quantum field theory}, Lecture Notes in Physics {\bf 51}, Singapore--New York--London--Hong Kong: World Scientific, 1993.
\bibitem{StrW} F. Strocchi, A. S. Wightman, Proof of the charge superselection rule in local, relativistic quantum field theory, J. Math. Phys. {\bf 15} 2198--2224 (1974).
\bibitem{Ta} M. Takesaki, Tomita's theorem of modular Hilbert algebras and its applications, Springer, Berlin, 1970.
\bibitem{Un} W. G. Unruh, Notes on black hole evaporation, Phys. Rev. D {\bf 14}, 870--892 (1976).
\bibitem{Yn} J. Yngvason, { On the algebra of test functions for field operators}, Commun. Math. Phys. {\bf 34}, 315--333 (1973)


\end{thebibliography}
\end{document}